\documentclass[jcp,aip,reprint]{revtex4-1}
\usepackage{bm,amsmath,epsfig}
\usepackage[version=3]{mhchem}
\usepackage{color}

\newcommand{\CR}{c_{\mathrm{R}}}

\newcommand{\nc}{n_{\mathrm{c}}}

\newcommand{\ee}{{\rm e}}

\newcommand{\eps}{\epsilon}

\begin{document}

\title{Optimising self-assembly through time-dependent interactions}
\author{Christopher J. Fullerton}
\email{christopher.fullerton@umontpellier.fr}
\affiliation{Department of Physics, University of Bath, Bath, BA2 7AY}
\affiliation{Laboratoire Charles Coulomb, UMR 5221, Universit\'e Montpellier, Montpellier, France}
\author{Robert L. Jack}
\affiliation{Department of Physics, University of Bath, Bath, BA2 7AY}

\begin{abstract}
We demonstrate a simple method by which time-dependent interactions can be exploited to improve self-assembly in colloidal systems. We apply this method to two systems: a model colloid with a short-ranged attractive potential, which undergoes crystallisation; and a schematic model of cluster growth. The method is based on initially strong bonds between particles, to accelerate nucleation, followed by a stage with weaker bonds, to promote growth of high-quality assembled structures. We track the growth of clusters during assembly, which reveals insight into effects of multiple nucleation events, and of competition between the growth of clusters with different properties.
\end{abstract}

\maketitle

\section{Introduction}
\label{sec:intro}

Self-assembly is the spontaneous formation of ordered states from simple components \cite{white02review,whitelam15-rev}. Examples include colloidal crysallisation~\cite{leunissen05, Hynn2007, sear07, chen11, romano11, klotsa-11}, viral capsid assembly \cite{zlotnick-2005,hagan-annrev}, and formation of tailored structures by DNA-mediated interactions~\cite{mirkin96, rothemund06, macfarlane11, nyk08, ke12}.
In many cases, the final assembled structure minimises the free energy of the system: this structure depends on the shape of component particles and the interactions between them \cite{sol07}.
Recent  developments in colloidal synthesis have allowed a high level of control over both of these elements \cite{nyk08, chen11, pawar09, sac10, jiang10, kraft12, sacanna13}, and both theoretical and simulation analyses have been made of the ordered phases that these units can form \cite{romano11, wilber09, akbari09, martinez11}.
However,  self-assembly is a dynamic process, and the successful design of assembly processes must consider the pathways between disordered and ordered states~\cite{whitelam15-rev}.
Often, if bonds between assembling components are strong, one finds long-lived disordered aggregates (kinetic traps) that disrupt self-assembly.
To avoid these effects, the formation of bonds must be microscopically reversible \cite{whitelam15-rev, hagan06, rap08, white02, whitelam09, grant-11}.
Frequent bonding and unbonding events allow errors that are made during assembly to be corrected before they get embedded in the bulk of the assembled object, at which point annealing of such errors (or defects) is very slow.

The twin requirements of a stable assembled structure and reversible bond formation lead to severe restrictions on the conditions that lead to effective self-assembly \cite{whitelam15-rev}.
Often, if bonds are weak enough to allow microscopically reversible growth, one finds a long induction time for assembly, due to the presence of a slow nucleation process.  In such cases, there is a tension between the best conditions for rapid nucleation of an ordered structure, and the best conditions for defect-free growth. This idea has a long history in crystal nucleation, as discussed by Galkin and Vekilov~\cite{vekilov99}, who used strongly supersaturated solutions of lysozyme protein to promote nucleation, followed by crystal growth under conditions of weaker supersaturation.  It has also been argued by Sch\"on that such a protocol is optimal for obtaining the best crystal yield in a finite time~\cite{schoen09}.  Klotsa and Jack~\cite{klotsa-13} proposed a rather general automated method for optimising the reversibility of self-assembly, which led to a similar protocol for time-dependent interaction strengths in colloidal crystallisation.

Here we consider time-dependent interactions in the context of colloidal self-assembly -- we focus on the example of  crystallisation, but we argue that similar results can be expected in other self-assembly processes too.  Our work is motivated by recent experiments such as those in  Ref.~\onlinecite{taylor12}, in which interactions between colloidal particles can be controlled in real time, and in which particle trajectories can be followed in detail during the crystallisation process.  Colloidal self-assembly is  different from molecular self-assembly and other nano-scale processes, in that the microscopic time scales for colloid motion are relatively slow (milliseconds to seconds), so there is a much weaker separation between these time scales and those which are experimentally accessible.  This offers new possibilities for exploiting time-dependent interactions in optimising self-assembly.

In the following, we present computer simulation results for colloidal crystallisation, and we also introduce a schematic model for cluster growth, which incorporates nucleation, growth, and kinetic trapping effects.  The models are defined in Sec.~\ref{sec:models} and an overview of results with fixed (time-independent) interactions is given in Sec.~\ref{sec:results-indpt}.  Then, in Sec.~\ref{sec:results-protocol}, we show results obtained for time-dependent interactions, showing how this can improve the yield of self-assembly processes, in both the models considered.  By tracking the clusters of particles that form during self-assembly, we elucidate the mechanism by which the time-dependent interactions improve the results.  In particular we draw an analogy between cluster growth and natural selection via survival of the fittest. We conclude in Sec.~\ref{sec:conc} and give an outlook as to future possibilities in this direction.  We also include two appendices, with Appendix~\ref{app:schem} collecting some additional analysis of the schematic assembly model that we introduce in this paper and Appendix.~\ref{app:clusters} describing details of the analysis of our simulation data.

\section{Models} \label{sec:models}

\subsection{Model colloidal system}

\begin{figure}
\begin{center}
\includegraphics[width = 8.5cm]{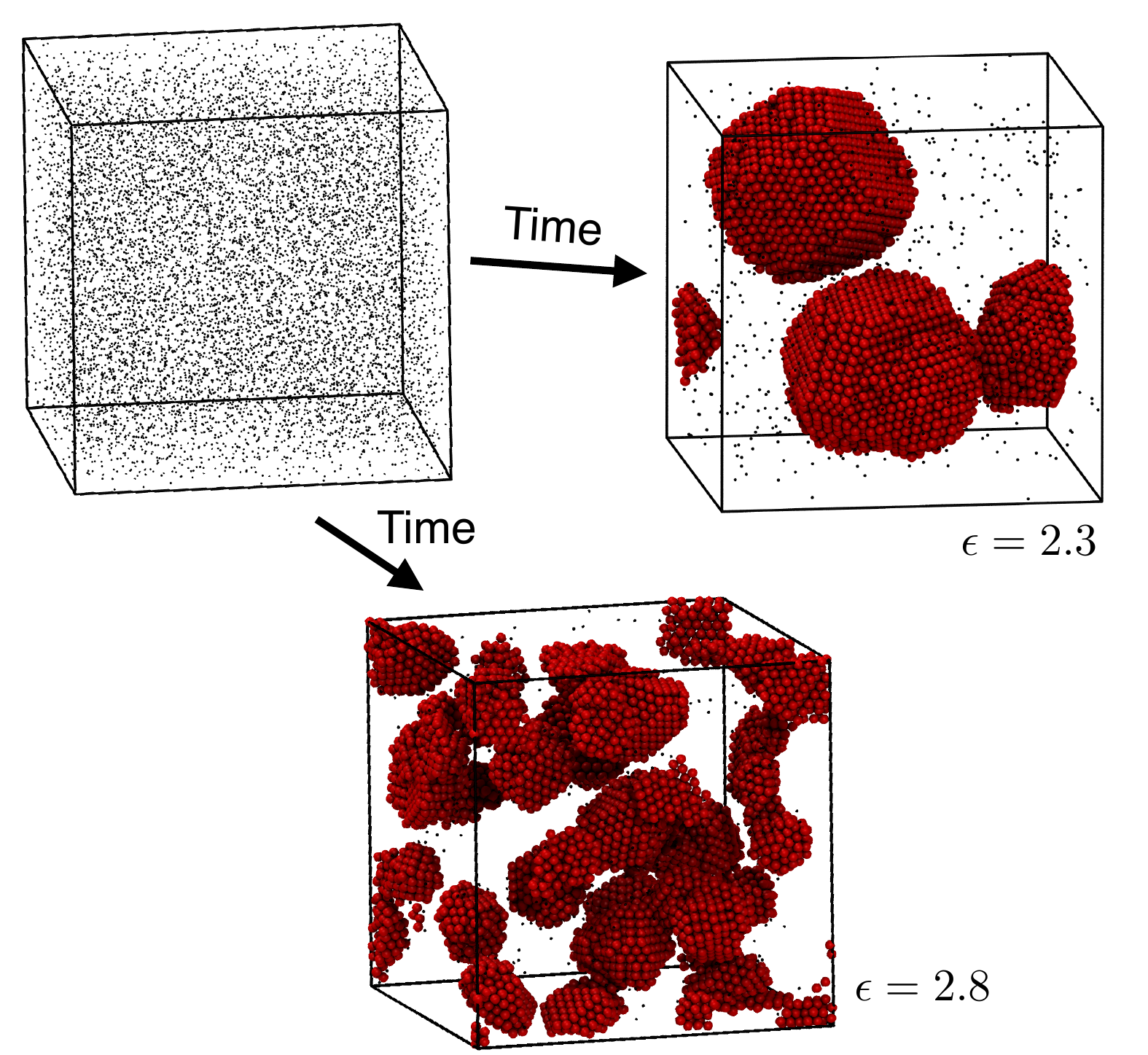}
\caption{%
Illustration of crystallisation in a model colloidal system~\cite{vmd}.  Starting from a disordered initial state in a system with a constant interaction strength $\eps$, one observes the formation of crystalline clusters, which are visualised at time $t=9100\tau_{\rm B}$.  Free particles (those which are not bonded to any others) are displayed at reduced size, for clarity.  The bond strength $\eps=2.3$ leads to large clusters with extended crystalline order.  For stronger bonds $\eps=2.8$, the clusters are smaller (due to multiple nucleation events) and kinetic trapping effects also mean that the degree of crystalline order is less (see Fig.~\ref{fig:qcompare}, below).
}
\label{fig:stickyspheres}
\end{center}
\end{figure}

As a model self-assembly process, we consider crystallisation in a system of of hard spherical particles that interact with each other through a square-well attractive potential.
This model mimics the behavior of colloid-polymer mixtures, in which small non-adsorbing polymers mediate attractive forces between colloids, via the depletion interaction~\cite{Poon02}.  Specifically, we consider $N=10000$ spherical particles of diameter $\sigma$ in a cubic box of side $L$, with periodic boundary conditions.  The box size is chosen so that the packing fraction $\phi=N\pi\sigma^3/(6L^3)=0.1$.  The particles interact by an interaction potential $U(r)$ which is infinite for $r<\sigma$; takes the value $-u$ for $\sigma<r<\sigma(1+\lambda)$; and is equal to zero for $r>\sigma(1+\lambda)$.  We take $\lambda=0.1$.   This square-well potential is a coarse approximation to a typical colloidal interaction potential, but it is sufficient to reproduce the qualitative features of these systems~\cite{klotsa-11, klotsa-13, haxton15}.  Indeed, for systems with short-ranged interactions, the observed behaviour typically depends very weakly on the precise form of the interaction potential~\cite{noro2000, rosenbaum96, rosenbaum99}.

To mimic the diffusive dynamics of the colloidal particles, we use a (single-particle) Monte Carlo (MC) method.  At each step, a particle is chosen at random, and given a random displacement taken from a cube of side $2a_0$.  Depending on the energy change associated with this displacement, the move is either accepted or rejected, according to the Metropolis criterion~\cite{frenkelsmit}.  Then, the time is $t$ incremented by $\tau_0=a_0^2/(6DN)$ where $D$ is the diffusion constant of a single colloidal particle.  In the limit of small $a_0$, this MC method is equivalent to solving an overdamped Langevin equation~\cite{whitelam11-molsim} [in the absence of interactions, the mean square displacement of a single colloid is $\langle r(t)^2 \rangle=6Dt$].  In practice we take $a_0=0.015\sigma$ -- this is significantly smaller than the range of the potential, but even this value is not yet small enough to be representative of the limit $a_0\to0$.  In particular, when large clusters of particles form (such as crystallites), the MC method tends to suppress the diffusion of these clusters.  To avoid such problems, collective move MC method might be used~\cite{whitelam11-molsim}, or perhaps a method that accounts for many-body hydrodynamic interactions.  However, previous work indicates that qualititative features of self-assembly are not strongly affected by the specific method used, so we retain the single-particle MC method, for simplicity.

Throughout this work, we take $\sigma=1$ as the unit of length.  The strength of the attractive interactions enters through the dimensionless parameter $\eps=u/k_{\rm B}T$.  Time is measured in units of the
Brownian time, $\tau_\mathrm{B}=\sigma^2/D$, in which time the mean squared displacement of a free particle is $6\sigma^2$. (That is, $\tau_{\rm B}$ is of the order of time taken for a particle to diffuse a distance of its diameter).  Clearly $\tau_\mathrm{B}=6N(\sigma/a_0)^2\tau_0$ so a single Brownian time corresponds to approximately $26700$ attempted Monte Carlo moves per particle.

For this model at volume fraction $\phi=0.1$, previous work has shown that the thermodynamic state of the system for $\eps\gtrsim 1.7$ consists of a close-packed crystal coexisting with a dilute colloidal fluid~\cite{fortini08,haxton15}.  However, starting from a homogeneous fluid and increasing the attraction strength $\eps$, the formation of this crystal is typically a very slow process, due to the large nucleation barrier.  In practice, crystallisation is observed in computer simulations only for $\eps\gtrsim 2.3$, which is close to the binodal line associated with a metastable liquid phase~\cite{fortini08,klotsa-11}.  For strong bonds $\eps\gtrsim 4.0$, kinetic trapping hinders effective crystallisation on the time scales accessible to simulation (and similar kinetic trapping effects are relevant in experiments too).  The behavior for near-optimal assembly is illustrated in Fig.~\ref{fig:stickyspheres}.  In the following, we will investigate how this crystallisation process can be facilitated by the use of time-dependent interactions between colloids.

\subsection{Schematic model of cluster growth}
\label{sec:schem-def}

\begin{figure}
\begin{center}
\includegraphics[width = 8.5cm]{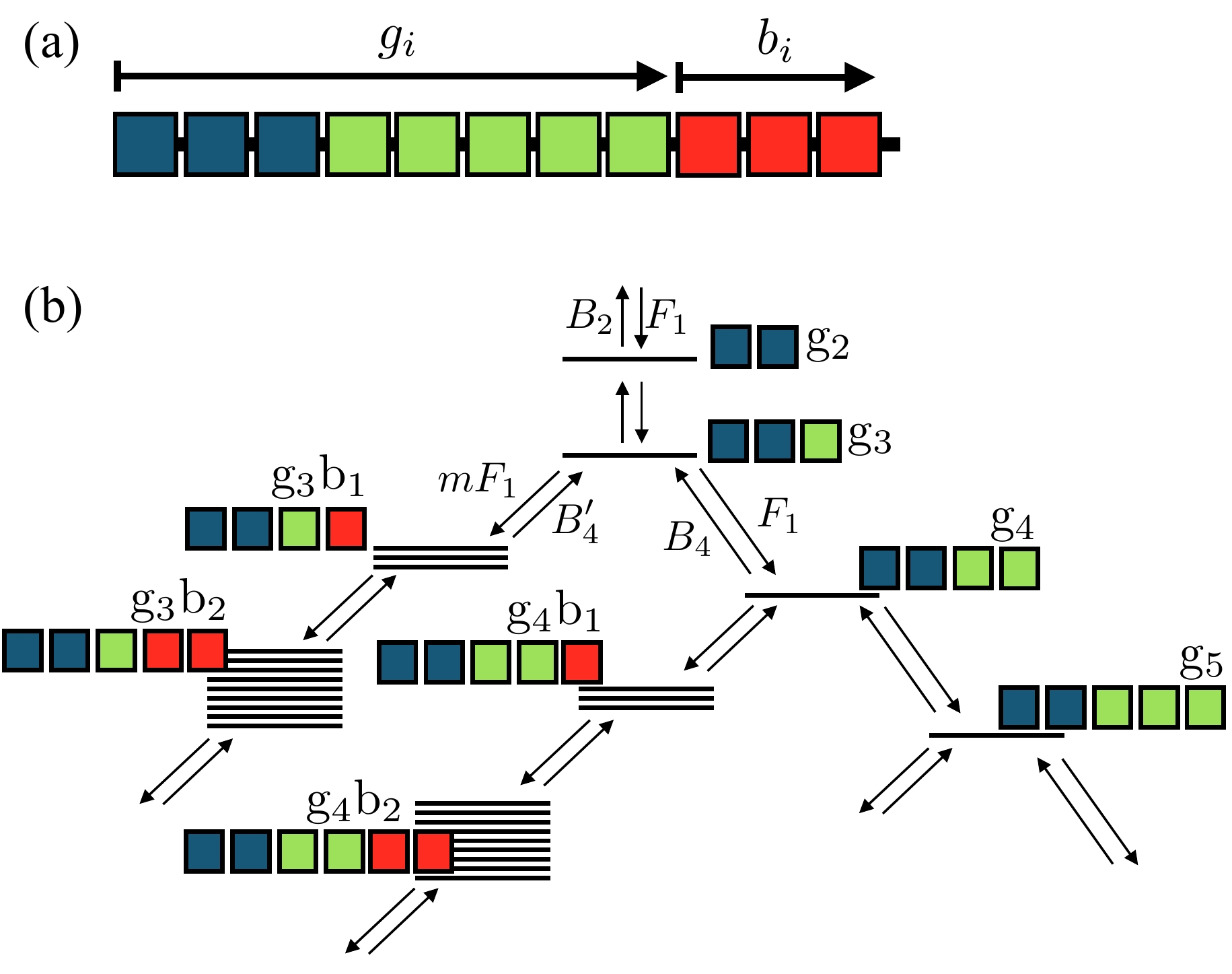}
\caption{%
Illustration of the schematic model of cluster growth.  
(a)~Each cluster can be viewed as a filament, which grows from left to right and contains up to three kinds of bonded particle.  The first $n_c$ particles (blue) are the nucleus, which is followed by some number of correctly bonded particles (green), and then a number of incorrectly bonded particles (red).  In this illustration $n_c=2$ although the numerical results in this work are all for the case $n_c=6$. The number of incorrectly bonded particles in filament $i$ is $b_i$; the number of correctly bonded particles (including the nucleus) is $g_i$.
(b)~Diagram illustrating growth and shrinkage of cluster species.  Each cluster species is represented as an energy level and selected rate constants for transitions between levels are indicated.  The energy levels for species that include $b$ incorrectly bonded particles have a degeneracy of $m^{b}$, reflecting the fact that each incorrectly bonded particle can bind in $m$ different ways.  (The illustrated case is $m=3$ but the results presented in this work are all for $m=4$.  See also Fig.~2 of Ref.~\onlinecite{grant-11}.)
}
\label{fig:filament}
\end{center}
\end{figure}

The second model that we consider here is a schematic model of growing clusters, that has been designed to capture the physics of nucleation, growth and kinetic trapping.  We use this model to analyse the effects of time-dependent interaction strength in a generic model of self-assembly, as evidence that the phenomena that we are investigating are potentially relevant for a variety of self-assembly processes.

The model builds on early work on phase transformation~\cite{becker35,binder76}, as well as recent analyses of viral capsid formation~\cite{hagan-10, zlotnick-2005} and amyloid fibrils~\cite{knowles2009}.  However, our model extends most previous studies because we explicitly incorporate kinetic trapping, by including the possibility that each particle in a cluster can either be \emph{correctly bonded} (that is, consistent with the final assembled structure that minimises the free energy) or \emph{kinetically trapped}.  A simple model that allows for these latter possibilities was discussed in Ref.~\onlinecite{grant-11}: here we incorporate similar features into a model that also includes both nucleation and growth.  A similar model involving growth and kinetic trapping (but not nucleation) was also recently considered by Whitelam, Dahal and Schmit~\cite{whitelam16-poison}.

Within the model, each cluster is  represented as a one-dimensional filament.  For a filament consisting of $i$ correctly bonded particles and $j$ kinetically trapped particles, we write a chemical formula for the cluster as $\ce{g}_i \ce{b}_j$, where $\ce{g,b}$ are shorthand notation for \emph{good} (correctly bonded) and \emph{bad} (trapped) particles.  Clusters of good particles (for example $\ce{g}_i$) can grow by the addition of either good or bad particles.  However, if a cluster includes any bad particles (for example $\ce{g}_i \ce{b}_j$), it can only grow by the addition of more bad particles.  The idea is that defective (bad) structures must be completely annealed before good growth can resume.

To be precise, we write chemical reaction equations for the filaments, 
\begin{align}
\ce{g}_i + \ce{g}_1 &
\ce{<=>[F_1][B_{i+1}]} 
\ce{g}_{i+1}
,  \nonumber \\
\ce{g}_i \ce{b}_j + \ce{g}_1 & 
\ce{<=>[mF_1][B_{i+1}']} \ce{g}_i \ce{b}_{j+1} . 
\label{equ:react}
\end{align}
Here $\ce{g}_1$ represents a free monomer.  The first reaction involves the formation of a correct (good) bond as the free monomer binds to a cluster.  The rate constant for the forward reaction is $F_1$; for the reverse reaction (filament shrinking) the rate depends (in general) on the size of the filament, and is denoted by $B_{i+1}$.  This reaction can take place for any value of $i$.  
(Recall that the definition of the rate constant means that the rate of change of concentration for clusters of size $i+1$ via the first reaction is $\partial_t c({\rm g}_{i+1}) = F_1 c(\mathrm{g}_1) c(\mathrm{g}_i)$ where $c(X)$ is the concentration of cluster $X$.)
Various choices for the size dependence of $B_{i+1}$ are possible~\cite{zlotnick-2005,knowles2009}: here we take a simple approach that separates pre-nucleation clusters $i\leq \nc$ from post-nucleation (growing) clusters $i> \nc$, details are given below.

The second reaction in (\ref{equ:react}) involves the formation of a kinetically trapped state (bad bond) as the free monomer binds.  The rate for this process is $m$ times larger than the rate for correct binding, reflecting the fact that there are typically many more ways to be kinetically trapped than there are to be correctly bonded. This reaction can take place for any value of $j$ (including $j=0$) but we assume that kinetic trapping occurs only in the post-nucleation (growth) phase of the self-assembly process, so this reaction can take place only for $i>\nc$: kinetic trapping is possible only after nucleation is complete and the first post-nucleation bond has formed.  For simplicity, we neglect processes such as fission or fusion of clusters (except via monomers), and secondary cluster nucleation~\cite{knowles2009}.

Having specified the forward rate constants as $F_1$ and $mF_1$, the backward rates are determined by free energy considerations.  On adding a correctly-bonded particle to a post-nucleation cluster, we suppose that the free energy is reduced by $u_0$.  For a pre-nucleation cluster, this free energy change is smaller, given by $u_0/\nu$ -- this reduction accounts for the fact that the free energy of small clusters is strongly affected by their surfaces, as in classical nucleation theory.  On adding an incorrectly bonded particle to a post-nucleation cluster, the free energy change is $(u_0/\mu)+k_{\rm B}T\log m$, where second term reflects the increased entropy associated with the kinetically trapped states.  As usual in such reaction schemes, these free energies are quoted at a reference concentration $c_{\rm R}$.  Denoting the free energy change by $\Delta \mathcal{F}$ and defining $\beta=1/(k_{\rm B}T)$, we note that the ratio of forward and backward rate constants is given (in the general case) by $B/F=\CR {\rm e}^{-\beta \Delta \mathcal{F}}$, leading to
\begin{align}
B_{i+1} &=  \begin{cases} F_1 \CR {\rm e}^{-\beta u_0}, & i\geq \nc \\ 
                                          F_1 \CR {\rm e}^{-\beta u_0/\nu}, &  i< \nc \end{cases} 
\\
B'_{i+1} & = F_1 \CR {\rm e}^{-\beta u_0/\mu}
\end{align}
It is useful to define $\eps=\beta u_0$ which is the dimensionless parameter that determines the strength of attractive interactions in this model.  All results shown here are for the case $m=4$, $\mu=4$, $\nu=5$, $\nc=6$.  This case is sufficient to illustrate the typical behaviour of the model, the main effects of these parameters are discussed in Appendix~\ref{app:schem}.  For the chosen parameters, we note that when bonds are strong enough to drive assembly, the free energy change for correct binding is larger than that for incorrect binding ($\beta u_0 > \beta u_0/\mu + \log m$), ensuring that large correctly-assembled are thermodynamically preferred to kinetically trapped state, as in the colloidal model.

In the equilbrium state, the average concentrations $\overline{c}$ of different species satisfy relations such as
\begin{equation}
\overline{c}(\mathrm{g_{i+1}}) = \frac{\overline{c}(\mathrm{g}_i) \overline{c}(\mathrm{g}_1)}{\CR} {\rm e}^{\beta u_0} , \qquad i\geq n_c,
\label{equ:eq-const}
\end{equation}
with similar equilibrium relationships for $i<n_c$ and for filaments including incorrect bonds: see Appendix~\ref{app:equ}.

We consider a total of $N$ particles in a system of volume $V$, such that the total particle concentration is $c_{\rm T}=N/V$.  However, the dependence of the system on the volume $V$ is encapsulated through the dimensionless parameter $c_{\rm T}/\CR$: all concentrations are measured in units of $\CR$ so the model is fully specified given the values of $(\eps,\nc,\mu,\nu,c_{\rm T}/\CR,N)$.  It is possible to take $\CR=1$ without any loss of generality, which corresponds to measuring all concentrations relative to $\CR$.  However, we retain $\CR$ in our equations so that all concentrations have units of inverse volume.  The unit of time is set by the rate constant $F_1$ and the reference concentration $\CR$, as $t_0=1/(F_1 \CR)$.

We take $c_{\rm T}/\CR=0.01$, which corresponds to a dilute system, as in the colloidal model.   Physically, note from (\ref{equ:eq-const}) that if $c_{\rm T}\approx \CR$ then the system will include large clusters even in the absence of attractive interations ($\eps=0$); working in the dilute case $c_{\rm T}\ll \CR$ ensures that assembly of clusters is driven by the attractive forces.  Moreover, since there are $N$ particles in total then one has a sum rule for the concentrations of clusters: 
\begin{equation}
\sum_X l(X) c(X)=c_{\rm T},
\label{equ:sum-rule}
\end{equation}
 where $l(X)$ is the number of monomers in species $X$ and the sum runs over all possible species.  

\section{Results -- interactions that are independent of time}
\label{sec:results-indpt}

\begin{figure}
\begin{center}
\includegraphics[width = 8cm]{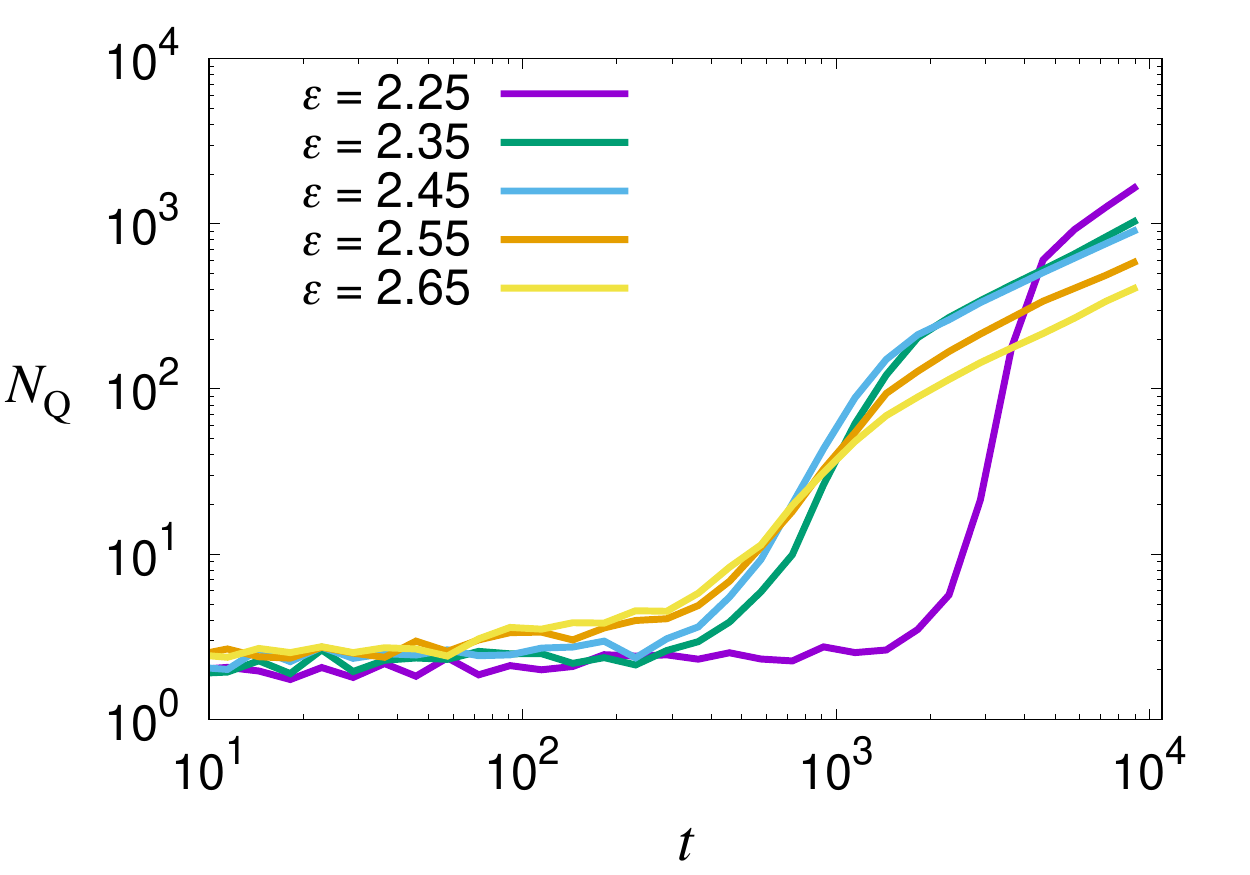}
\caption{Time dependence of the crystallinity order parameter $N_Q$ in the model colloid, for various bond strengths $\eps$.  Time is measured in units of the Brownian time $\tau_{\rm B}$.  For $\eps=2.25$ there is a substantial lag time before nucleation takes place.  For $\eps\geq2.35$ the lag time is almost independent of bond strength, indicating that nucleation is not a rare event: in this case the formation and growth of clusters is controlled by kinetic parameters related to particle collisions.}
\label{fig:qcompare}
\end{center}
\end{figure}

\begin{figure}
\begin{center}
\includegraphics[width = 8cm]{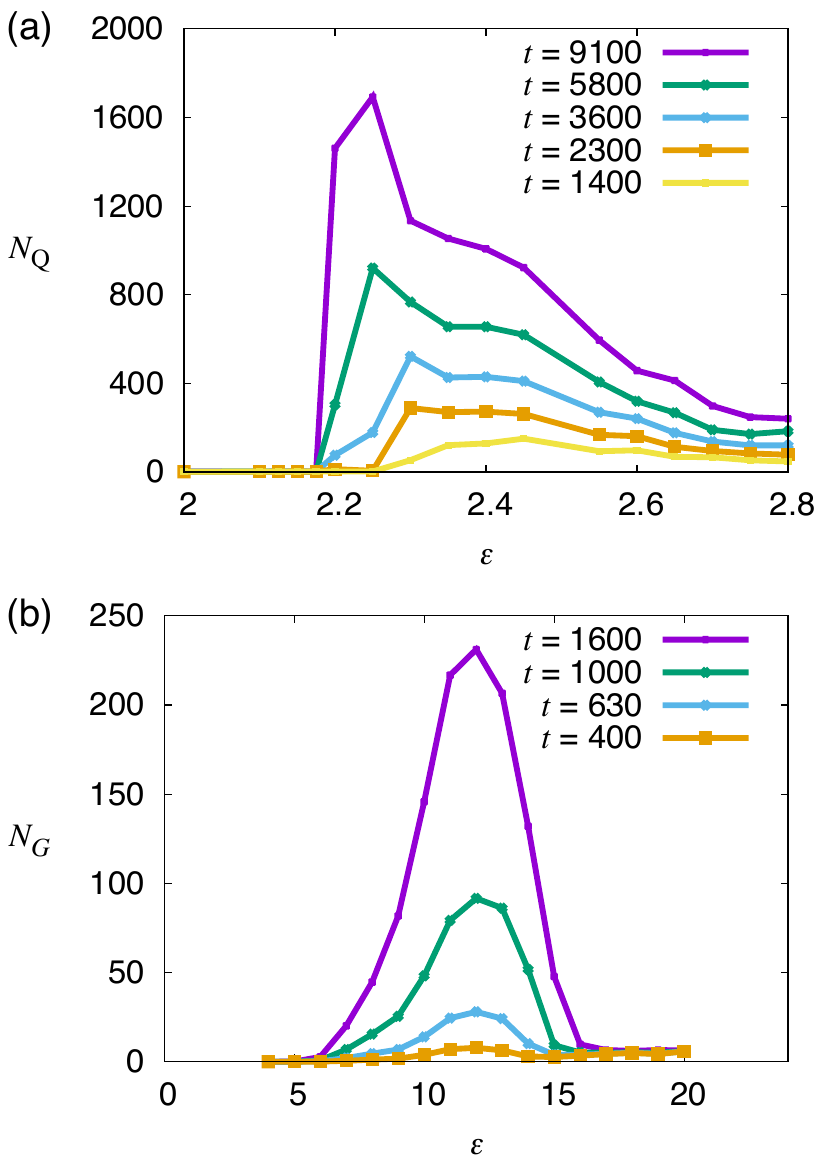}
\caption{Measures of yield for self-assembly as a function of bond strength $\eps$ at various times $t$.  (a) Model colloid. The measure of crystallinity $N_Q$ shows a non-monotonic dependence on $\eps$, with no crystals observed when bonds are weak (due to the typical nucleation time being much larger than $t$) and low crystallinity when bonds are strong (due to kinetic trapping).
b) Schematic model.  The measure of good-quality assembly $N_G$ shows similar non-monotonic dependence on $\eps$.}
\label{fig:yieldconste}
\end{center}
\end{figure}

The main focus of this work is the behaviour of the colloidal and schematic models when the strength of particle interactions $\eps$ depends on time.  In this section we consider interactions that do not depend on time, as a baseline for the case of time-dependent interactions considered in Sec.~\ref{sec:results-protocol}, below.

\subsection{Model colloidal system}

In order to assess the extent to which a crystallisation (or self-assembly) process is effective, it is important to define a measure of the quality of the assembled product.  In the colloidal system, we note that perfect crystals exhibit long-ranged translational and bond-orientational order, but local measurements of particle environments are not sufficient to assess whether such long-ranged correlations have formed. (For example, so-called nanocrystalline states~\cite{Mithen16} can have crystalline local packing but negligible long-ranged order.)  To define an order parameter that is sensitive to long-ranged crystalline order, we use a global measure of bond-orientational correlations, calculated as follows~\cite{fortini08, tenwolde95, klotsa-13}.
For a configuration of the system, the bonds particle $p$ makes with its neighbours are projected onto the spherical harmonics with $l = 6$ to give a complex vector $\vec{q}_6(p)$, normalised so that $\vec{q}_6(p) \cdot \vec{q}_6(p)^*=1$ (the asterisk denotes complex conjugation).
The vectors $\vec{q}_6$ are summed over all particles to give $\vec{Q} = \sum_{p=1}^{N}\vec{q}_6(p)$ and finally $N_Q$ is given by
\begin{equation}
N_Q = N^{-1}\langle\vec{Q}^* \cdot\vec{Q}\rangle.
\label{equ:NQ}
\end{equation}
As discussed in Ref.~\onlinecite{klotsa-13}, $N_Q$ can be interpreted as a measure of the crystalline domain size in the system -- if all particles are arranged in a perfect crystal, then $N_Q$  takes its maximal possible value of $N$.  Since $N_Q$ measures domain size, it can also distinguish systems with two small crystallites from those with one larger crystalline cluster -- this distinction is important for assessing the quality of the assembled crystals, but local measurements of structure cannot distinguish these cases.  

We simulated the self-assembly of the model colloidal system, starting from an initial state with hard particles distributed at random, and running the MC dynamics.  Fig.~\ref{fig:qcompare} shows how the crystallinity parameter $N_Q$ increases with time during crystallisation, over a narrow range of interaction strengths.  For $\eps=2.25$, one finds $N_Q\approx 2000$, significantly smaller than the total number of particles $N=10000$.  This result is consistent with Fig.~\ref{fig:stickyspheres}, which shows that multiple crystalline clusters tend to form under these conditions, due to multiple nucleation events. This effect reduces the typical domain size.  Note that the MC dynamical method used here may underestimate the extent to which these clusters would diffuse and collide, leading to a large cluster that consists of multiple domains and therefore lacks long-ranged bond-orientational order.  However, such processes have little effect on the value of $N_Q$ so we expect the main results presented here to be robust even if diffusion of large clusters was included (for example by a collective move MC method~\cite{whitelam11-molsim}).

\subsection{Schematic model -- assembly yield}

We also simulated cluster growth (that is, self-assembly) in the schematic model, starting from an initial condition in which all particles are free monomers.  
To define a measure of assembly yield, one candidate would be the total number of particles in correctly bonded environments.  However, as in the colloidal model, this does not distinguish configurations with many small clusters from correctly-assembled states, which have \emph{small numbers} of \emph{large clusters}.  To make this distinction, we require an analogue of $N_Q$ in this schematic model.  To achieve this, suppose that we choose a particle at random, and we measure the number of correctly bonded particles in the cluster that contains that particle.  If $g_f$ is the number of good particles in cluster $f$ and similarly $b_f$ is the number of bad particles then the average cluster quality defined in this way is
\begin{equation}
N_G = N^{-1} \left\langle \sum_f g_f ( g_f + b_f ) \right\rangle
\end{equation}
where the sum runs over all clusters, except for free monomers (hence $N_G=0$ at time $t=0$).

\subsection{Assembly yield measurements}

Fig.~\ref{fig:yieldconste} shows  measurements of $N_Q$ and $N_G$ in the colloidal system and the schematic cluster model, as a function of the bond strength.  We interpret these results as measures of the \emph{yield} of the self-assembly processes. For each bond strength, several independent simulations were performed: we show average values of $N_Q$ and $N_G$ for various times.  (Averages are taken over 16 trajectories in the model colloid and around 1000 trajectories in the schematic model.) The behavior shown in Fig.~\ref{fig:yieldconste} follows the expected form in such models~\cite{hagan06, klotsa-11, whitelam09, grant-11, jankowski11}: when $\eps$ is small, nucleation is very slow and no assembly is found; when $\eps$ is large kinetic trapping effects lead to less effective self-assembly (due to growth of disordered clusters and multiple nucleation effects).  There is a narrow range of bond strengths $\eps$ in which assembly leads to large crystalline (or correctly-bonded) clusters.  Optimising experimental conditions in order to find this narrow range of parameter values is a difficult and practically-important task.  In the following, we demonstrate how this problem might be avoided by exploiting time-dependent interactions.

\subsection{Schematic model -- further analysis}
\label{subsec:shemactic-model-further-analysis}

Our main focus here is on the model colloidal system, and we use the schematic model below to illustrate general features of assembly with time-dependent interactions.  However, the schematic model itself has a rich phenomenology, even with interactions that are independent of time.  We defer a full analysis of these effects to a later work (see also Ref.~\onlinecite{whitelam16-poison}), but we include in Appendix~\ref{app:schem} an overview of the relevant behaviour, in order to set the present results in context.  This subsection summarises those results.

The schematic model encapsulates the physics of nucleation, growth, and kinetic trapping.  Significant clusters in the system form for $\eps\gtrsim \eps^*$ with $\eps^*\approx \ln(\CR/c_{\rm T})$.  For the parameters considered here $\eps^*\approx 5$.  At equilibrium, correctly assembled clusters are preferred over kinetically-trapped ones for $\eps(\mu-1) > \mu\ln m$: for the parameters considered here, this is satisfied whenever $\eps\gtrsim \eps^*$.  However, incorrectly-bonded clusters tend to grow at the expense of correctly-bonded ones whenever $c_0 > c_{\rm R} \ee^{-\eps/\mu}$, which means that kinetic trapping is relevant for $\eps\gtrsim \eps^{\rm trap}$ with $\eps^{\rm trap}\approx \mu \ln (\CR/mc_{\rm T})$.  For the parameters considered here $\eps^{\rm trap}\approx 14$.

\begin{figure}
\begin{center}
\includegraphics{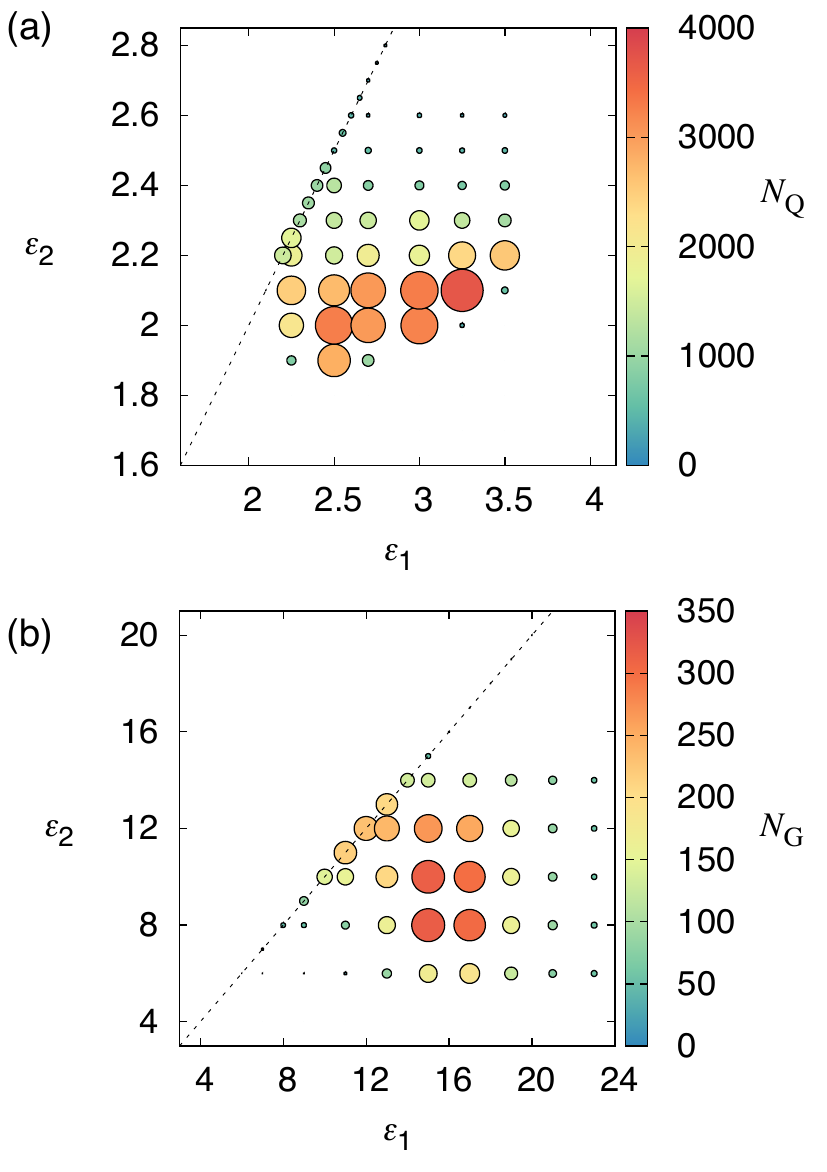}
\caption{Measures of yield for self-assembly with time-dependent interactions.  The bond strength is $\eps_1$ for $t<t_{\rm step}$ and $\eps_2$ for $t\geq t_{\rm step}$.  Colored circles indicate the measure of assembly quality ($N_Q$ or $N_G$) at the final time $t_{\rm end}$, with the sizes of the circles also proportional to the value of the relevant observable.  Dashed lines indicate the cases with constant interaction strengths $\eps_1=\eps_2$.  
(a)~Model colloid, with $t_{\rm step}=2300\tau_{\rm B}$ and $t_{\rm end}=9100\tau_{\rm B}$. (b)~Schematic model with $t_{\rm step}=40\tau_{\rm 0}$ and $t_{\rm end}=10^4\tau_{0}$.
}
\label{fig:yieldvare}
\end{center}
\end{figure}

For relatively weak bonds, the system also supports a metastable state, from which a nucleation process must take place before clusters can grow.  This metastable state is relevant for $\eps^* < \eps \lesssim \eps^{\rm qe}$ with $\eps^{\rm qe}=\nu [ \ln (\CR/c_{\rm T}) - (\ln N)/(n_c-1) ]$, for stronger bonds $\eps\gtrsim \eps^{\rm qe}$ nucleation is no longer a rare event and there is no metastable pre-nucleation state.  For the parameters chosen here, $\eps^{\rm eq} \approx 14$, comparable with $\eps^{\rm trap}$.  We attribute the position of the peak of the assembly yield in Fig.~\ref{fig:yieldvare}(b) to a combination of kinetic trapping and multiple nucleation events, which dominate the system for $\eps\gtrsim 14$.

\begin{figure}
\begin{center}
\includegraphics[width = 8cm]{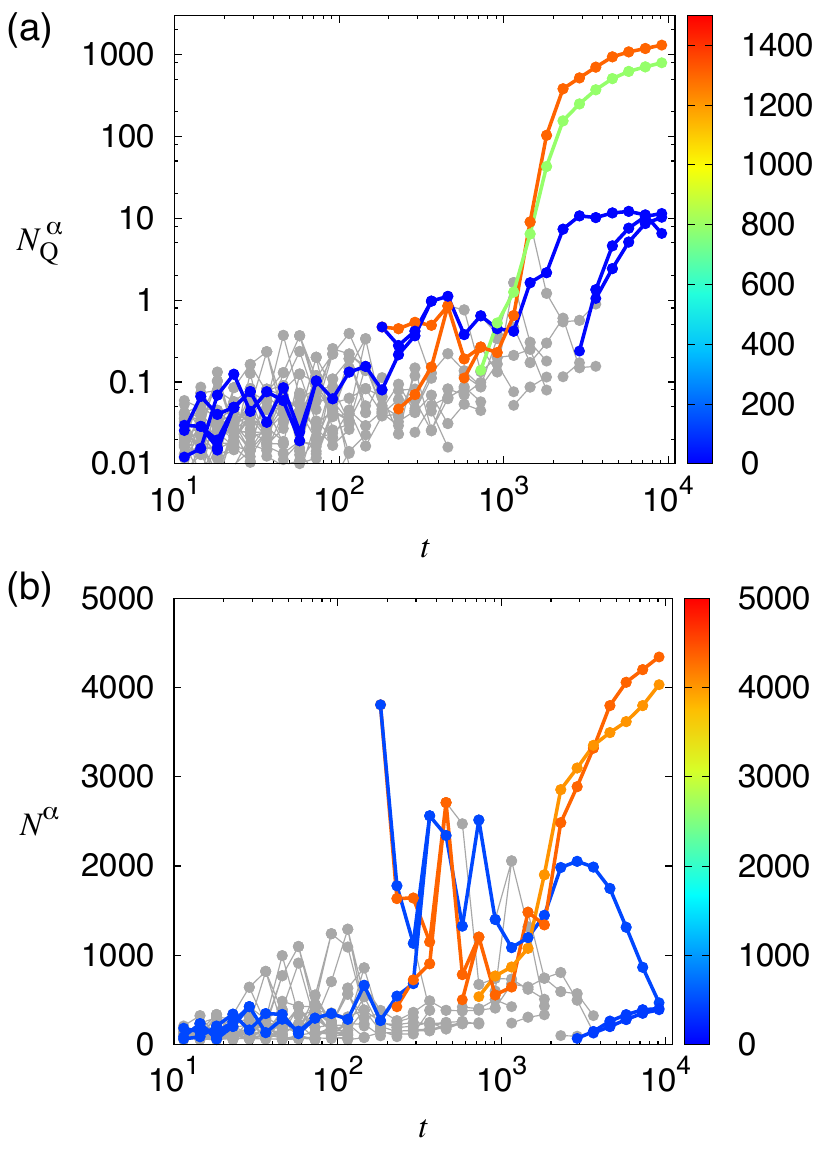}
\caption{Time series for a representative trajectory of the model colloid for fixed interaction strength $\eps=2.3$ (near-optimal assembly), illustating how clusters of particles nucleate, grow, and shrink.    (a)~Cluster crystallinities $N_Q^\alpha$ as a function of time, shown with a logarithmic vertical scale.  The lines are color-coded according to the final value of $N_Q^\alpha$, with grey lines and points for clusters that do not survive until the final time $t_{\rm end}$.  
Two large clusters appear at time $t\approx 10^3\tau_{\rm B}$ and grow quickly.  Smaller crystalline clusters are also apparent.
(b)~Cluster sizes $N^\alpha$ for the same trajectory, with a linear vertical scale.  Large clusters ($N^\alpha\approx 1000$) exist in the system for times $t\gtrsim 50\tau_{\rm B}$ but comparison with panel (a) shows that these are non-crystalline and do not grow rapidly until the formation of substantially crystalline clusters at times $t\approx 10^3\tau_{\rm B}$.  Some large clusters also shrink with time.
}
\label{fig:treeconste}
\end{center}
\end{figure}

\section{Results -- time-dependent interactions}
\label{sec:results-protocol}

\subsection{Assembly yield}

As discussed in Sec.~\ref{sec:intro}, our choice of time-dependent interactions is motivated by the idea that the optimal conditions for nucleation are not the same as those for crystal growth.  To address this problem in a simple way,
we consider simulation protocols  where the bond strength $\eps=\eps_1$ is fixed during a time period $0<t<t_{\rm step}$.  At time $t_{\rm step}$, the bond strength is reduced to $\eps_2$ and the self-assembly simulation is continued.  At time $t_{\rm end}$, the yield ($N_Q$ or $N_G$) is measured.  

For the model colloid we take $t_{\rm end}=9100\tau_{\rm B}$ (sufficiently long to observe significant crystallisation and consistent with accessible time scales in experiments~\cite{taylor12}).  We take $t_{\rm step}\approx t_{\rm end}/4$ which significantly improves the yield of the self-assembly process.  However, the behavior shown here is robust over a range of $t_{\rm step},t_{\rm end}$.  For the schematic model we take $t_{\rm end}=10^4\tau_0$ with the interactions changing at a rather earlier time $t_{\rm step}=40\tau_0$.  
Comparing with the model colloid, the time $t_{\rm step}$ is significantly smaller: the reason is that if multiple correctly-bonded clusters have nucleated in this model, there is no Ostwald ripening effect whereby the large ones grow at the expense of the small ones (see appendix \ref{app:quasi-eq-pre-nuc}).  This means that suppression of multiple nucleation events is particularly important for achieving large assembly yields, and a small value of $t_{\rm step}$ helps to achieve this.

Results are shown in Fig.~\ref{fig:yieldvare}.
Comparing the behaviour for $\eps_1\neq\eps_2$ with the case of fixed interaction strength ($\eps_1=\eps_2$, dashed line), the yield of both processes is significantly improved by the use of time-dependent interactions.  The physical idea is that the best conditions for crystal growth are rather different from the conditions for fast nucleation.  In particular, for bond strengths $\eps<2.1$ in the colloidal systems, nucleation is not observable on these time scales, so stronger bond than this are required to promote nucleation.  However, strong bonds tend to promote  kinetic trapping, so it is convenient to reduce the bond strength in order to promote growth of correctly assembled crystals.  Also, strong bonds at short times can promote multiple nucleation events -- weakening the bonds at later times typically causes some of the resulting clusters to shrink and vanish, allowing large good-quality crystals to grow.  This can be interpreted as an acceleration of Ostwald ripening, as we now discuss.

\begin{figure}
\begin{center}
\includegraphics[width = 8cm]{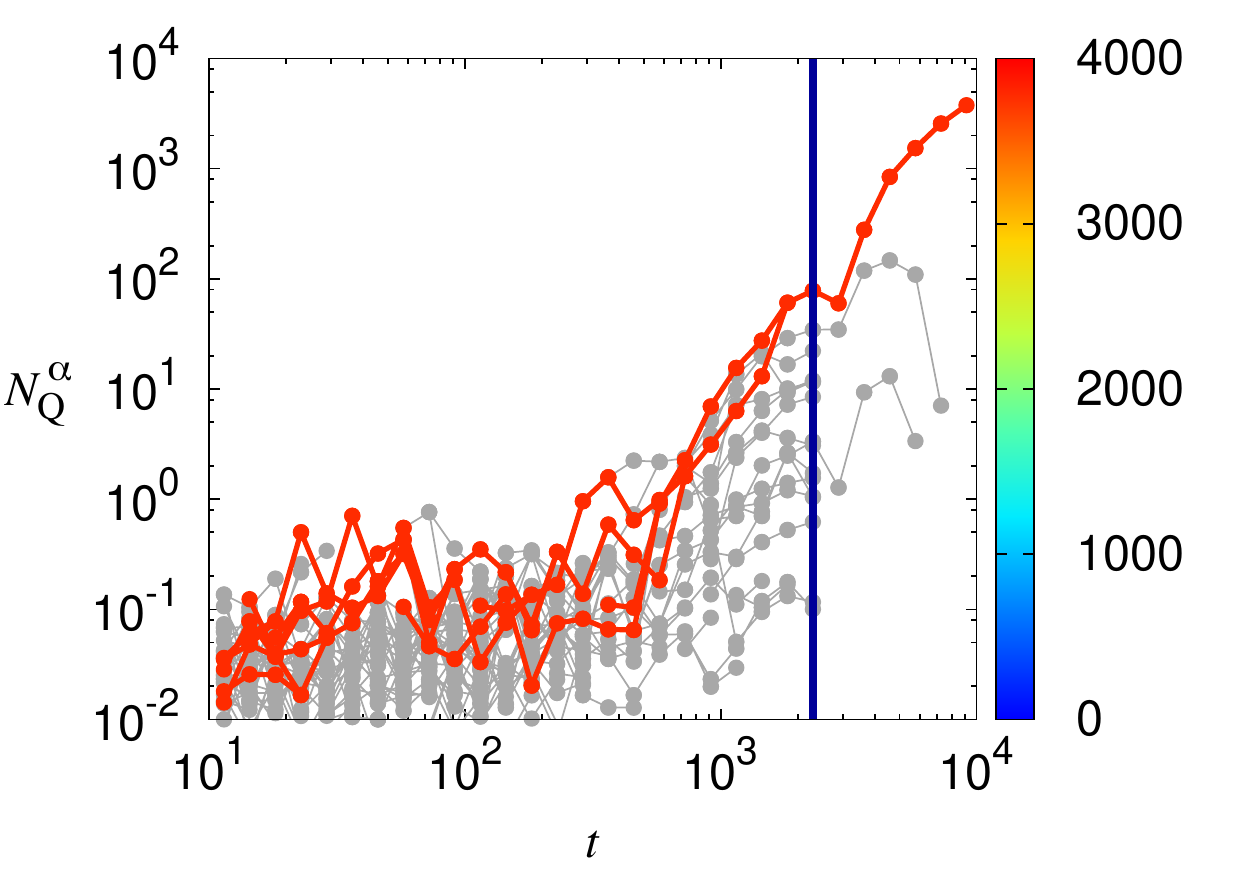}
\caption{%
Time series for a representative trajectory of the model colloid where the bond strength is
is initially $\epsilon_1 = 2.5$ and is changed to $\epsilon_{\mathrm{B}}^2 = 1.9$ at $t_{\mathrm{step}} = 2300\tau_{\rm B}$.  (The time $t_{\rm step}$ is illustrated with a vertical line.)
Comparison with Fig.~\ref{fig:yieldconste} shows that for $t<t_{\rm step}$ there are many clusters with significant crystallinity, but for $t>t_{\rm step}$ most of these clusters shrink and vanish, allowing a single large crystallite to grow.
}
\label{fig:treevare}
\end{center}
\end{figure}

\begin{figure}
\includegraphics[width = 6cm]{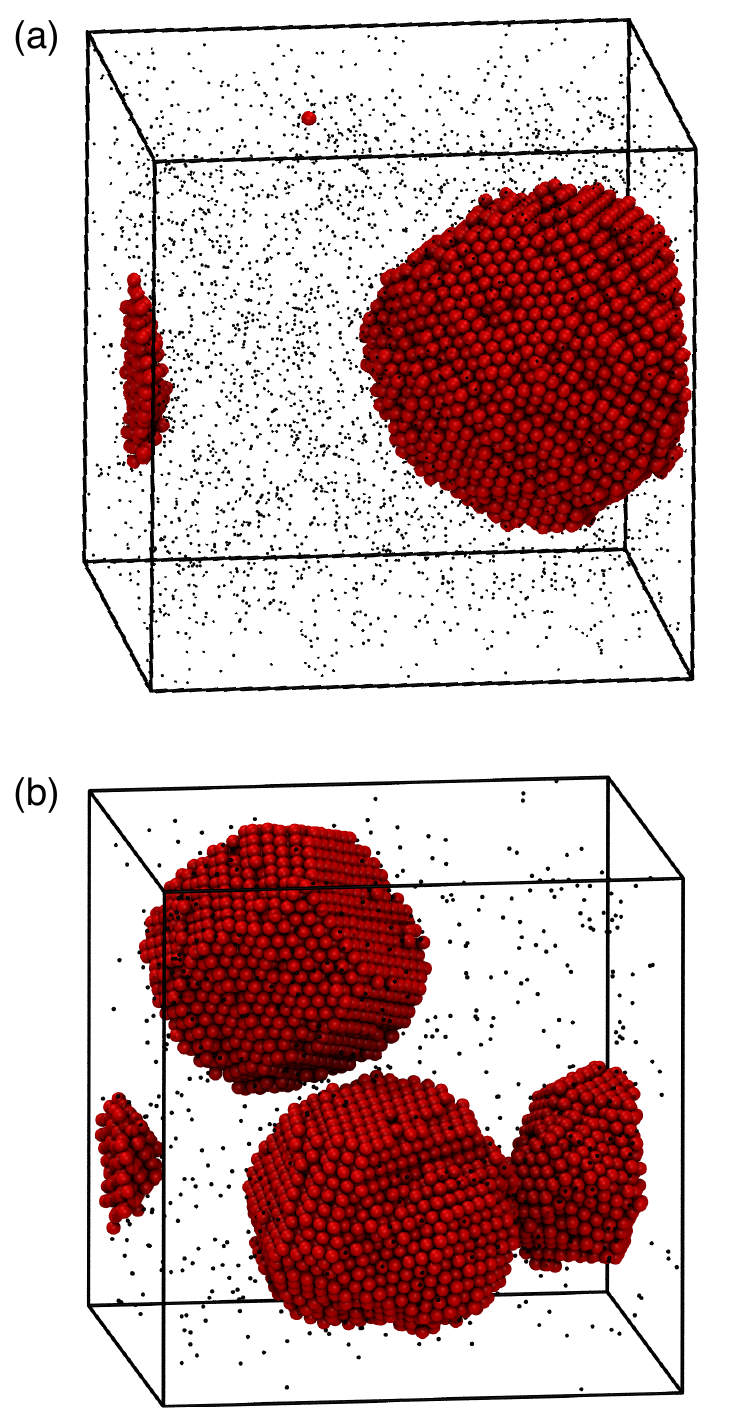}
\caption{
Representative snapshots of the model colloid at $t=t_{\rm end}=9100\tau_{\rm B}$.
(a) Time-dependent interactions with $(\eps_1,\eps_2)=(2.5,1.9)$ and $t_{\rm step}=2300\tau_{\rm B}$.
(b) Fixed interaction strength with the near-optimal value $\eps = 2.35$.  The time-dependent interactions reduce effects of multiple nucleation and kinetic trapping which reduce the crystallinity in the system with fixed interaction strength.
}
\label{fig:snapshot}
\end{figure}

\subsection{Tracking clusters during assembly}
\label{subsec:trackclusters}

To shed light on what makes assembly successful under certain conditions we track clusters in the colloidal system as they nucleate, grow and shrink with time.
A cluster is defined as a set of bonded particles.  For each cluster $\alpha$, let $N^\alpha$ be the number of particles in that cluster.  Also we define $N_Q^\alpha$ as a measure of the cluster crystallinity, by defining $\vec{Q}_\alpha=\sum_{p\in\alpha} \vec{q}_6(p)$ as a sum over particles in the cluster and replacing $Q\to Q_\alpha$ in (\ref{equ:NQ}) so that $N_Q^\alpha=N^{-1} \vec{Q}^*_\alpha \cdot \vec{Q}_\alpha$. (There is no average in this definition since we consider a single cluster. Note also that the normalisation factor is the total number of particles $N$, so $N_Q^\alpha$ is not itself a domain size: assuming that bond order parameters of each cluster are indepedent then the average domain size for the whole system $N_Q$ is obtained by summing $N_Q^\alpha$ over all clusters.)

Fig.~\ref{fig:treeconste} shows the sizes $N^\alpha$ and crystallinities $N_Q^\alpha$ of clusters in a single trajectory, as a function of time, during crystallisation with a constant bond strength $\eps=2.3$.  Clusters that are connected by growth, fusion or fission are connected with lines.
However, given a simulation trajectory containing configurations at various times $t_1,t_2,\dots$, it is not trivial to identify how clusters in different configurations are connected to each other.  Clusters can exchange particles between one another, they can merge or split, and new clusters can be created by nucleation.  In Appendix~\ref{app:clusters}, we describe the method that we used to follow the time evolution of clusters within the system.  This method determines how the points in Fig.~\ref{fig:treeconste} are connected to each other, to indicate their evolution as a function of time.  The method is not perfect -- one should assume that some clusters which are not connected in Fig.~\ref{fig:treeconste} are related by merging or splitting events, but we argue that the figure does illustrate the main points of interest.

In the classical picture of nucleation, growth and Ostwald ripening, one expects clusters to appear after a nucleation lag; they should grow until the free particles in the system are exhausted; and finally exchange of particles between clusters should lead to growth of large clusters and suppression of smaller ones.  Fig.~\ref{fig:treeconste} is approximately consistent with this picture, except that it is clear from Fig.~\ref{fig:treeconste}b that the Ostwald ripening regime does not always involve the growth of large clusters at the expense of small ones; instead Fig.~\ref{fig:treeconste}a shows that crystalline clusters with large $N_Q^\alpha$ tend to grow at the expense of less crystalline clusters (with smaller $N_Q^\alpha$).  

\begin{figure}
\begin{center}
\includegraphics[width = 8cm]{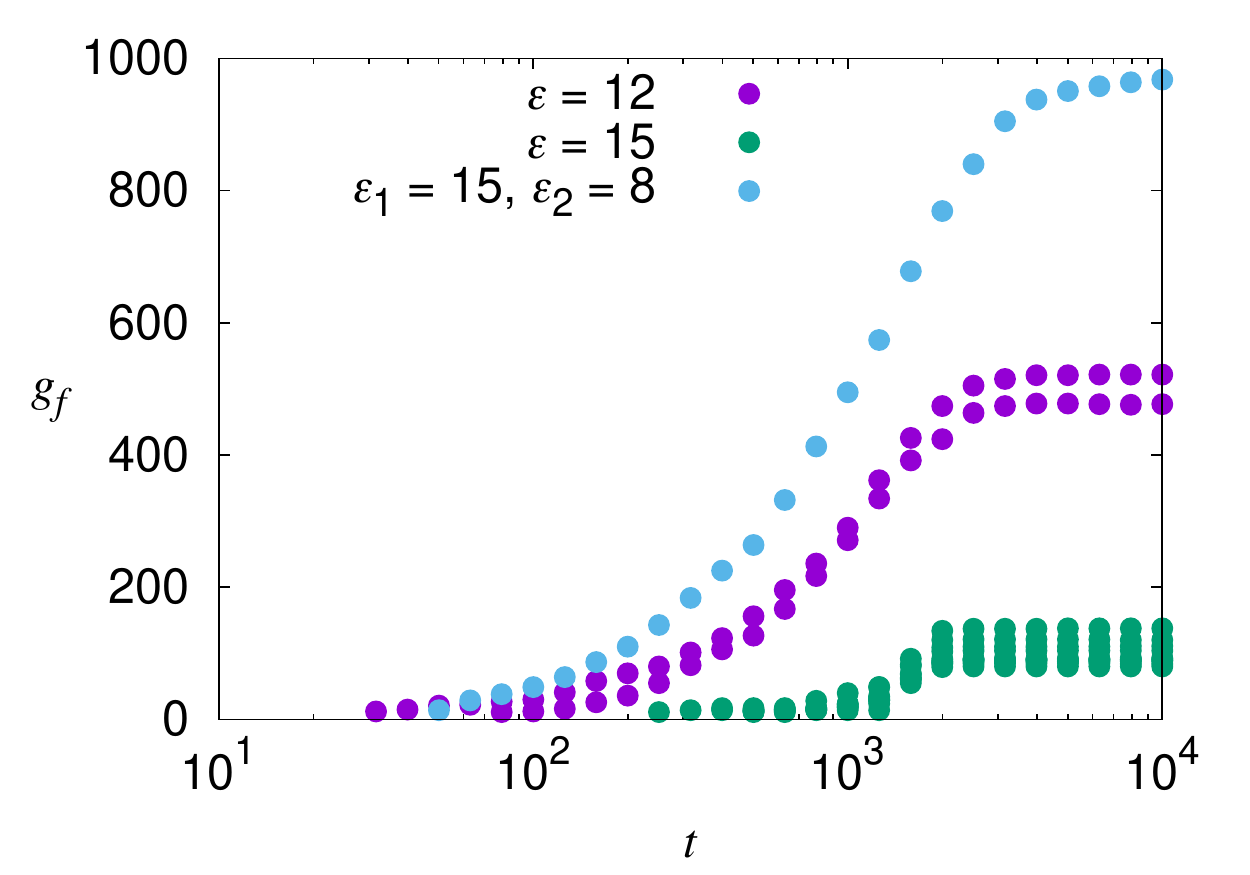}
\caption{%
Results showing growing clusters in the schematic model.  We plot the number of correctly-bonded particles $g_f$ for growing clusters within three different trajectories. Two of the trajectories involve with fixed interaction strength $\eps=(12,15)$ and one with time-dependent strength $(\eps_1,\eps_2)=(15,8)$ and $t_{\rm step}=40\tau_0$.  For fixed bond strengths one observes multiple large clusters, which limit the total size to which they grow, but the use of time-dependent interactions reduces this effect.  These results can be compared with Figs.~\ref{fig:treeconste},\ref{fig:treevare} in which case $g_f$ is analogous to $N_Q^\alpha$, but note that in this Figure we show results from three different trajectories, with the color indicating the trajectory: this contrasts with Figs.~\ref{fig:treeconste},\ref{fig:treevare}, each of which shows just one trajectory.
}
\label{fig:fil_treevare}
\end{center}
\end{figure}

To illustrate the effect of time-dependent interactions, consider Fig.~\ref{fig:treevare}.  Comparing with Fig.~\ref{fig:treeconste}, the initial bond strength is higher, so there are multiple nucleation events, leading to many clusters with a range of values of $N_Q^\alpha$.  On reducing the bond strength at time $t_{\rm step}$, the smaller and less crystalline clusters all shrink and vanish almost immediately, allowing the most crystalline cluster to grow.  At the end of the simulation, only one large crystalline cluster remains.  The result is illustrated in Fig.~\ref{fig:snapshot}a, in which one sees a single large crystalline cluster, compared with the effects of multiple nucleation seen for a constant interaction strength $\eps=2.35$ in Fig.~\ref{fig:snapshot}b.

Finally, Fig.~\ref{fig:fil_treevare} shows the behaviour of growing clusters in the filament model.
{At $\eps = 12$, the nucleation time is long but eventually two filaments nucleate and grow to roughly equal size. At a higher binding energy of $\eps = 15$, nucleation occurs much more quickly, but the clusters tend to form incorrect bonds and the system gets kinetically trapped.  For times $t \sim 10^3\tau_0$, the system escapes the trap, and correctly-bonded clusters grow.  However, there are many such clusters and this limits their maximum size.  Time-dependent interactions can be used to avoid the problems caused by slow nucleation, multiple nucleation events and kinetic trapping.  An example of this effect is also shown in Fig. ~\ref{fig:fil_treevare} -- for early times then we take $\eps_1 = 15$ so that the nucleation time is fairly short; then  at $t_{\rm step} = 40\tau_0$ the interaction strength is changed to $\eps_2 = 8$. This change in bond strength is at an early enough time that only a few clusters have nucleated, and it means that growth of kinetically trapped clusters is no longer favourable. The result is that one cluster grows until it contains nearly all of the monomers in the system.}

\subsection{Survival of the fittest, and Ostwald ripening}

To interpret the cluster tracking diagrams in Figs.~\ref{fig:treeconste},\ref{fig:treevare},\ref{fig:fil_treevare}, we invoke an analogy with survival of the fittest and natural selection.  The nucleation process in these systems leads to the formation of a population of clusters, which vary in their size $N^\alpha$ and their crystallinity $N_Q^\alpha$. (In the schematic model, the analogue of $N_Q^\alpha$ is the number of good particles in the cluster $g_f$).  After nucleation, these clusters grow quickly until the population of free monomers is almost exhausted, at which point cluster growth becomes slow, and one enters the Ostwald ripening regime.

In this regime, the classical picture is that particles are constantly binding and unbinding from the surface of clusters.  For spherical homogeneous clusters, unbinding happens more quickly when the clusters are small, since the curvature of their boundaries reduces the binding energy of surface particles.  This leads to a gradual growth of larger clusters, and shrinkage of smaller ones.  In our analogy, we interpret the  free monomers as a resource for which the clusters are competing.  Larger clusters are more effective in holding onto this resource, due to their reduced surface curvature.  We interpret this effect as an improved fitness for the large clusters, which means that they grow at the expense of the smaller ones.  

Since the cluster size determines whether it tends to grow or shrink, there is also an analogy between the cluster fitness and a reaction co-ordinate for cluster growth: the larger or more crystalline is a cluster, the more likely it is to grow, similarly to the situation in classical nucleation theory~\cite{whitelam15-rev} but now in the growth regime.

In the crystal-forming systems considered here, Fig.~\ref{fig:treeconste}b shows that the fitness of a cluster is not simply given by its size, since large clusters often shrink and small clusters often grow.  However, the crystallinity measure $N_Q^\alpha$ provides a better indication of the fate of a cluster: one can identify fitter clusters as those which are both larger and more crystalline.  That is, the existence of defects or structural disorder within clusters limits their ability to absorb monomers from the system, so such clusters tend to shrink as a function to time, at the expense of high-quality crystals.

With this in mind, we return to Fig.~\ref{fig:treevare}, in which the bond strength is reduced at time $t_{\rm step}$.  At this time, unbinding of monomers from clusters gets more likely.  We can interpret this as a reduction in fitness for all clusters or, more usefully, as an increase in the selection pressure in their environment.  As a result, clusters with lower fitness tend to shrink rapidly and vanish, allowing the  the fittest cluster to grow and absorb all of the available resources from the environment.  This is the mechanism by which a single high-quality crystalline cluster can grow within the system, leading to self-assembly with improved yield.   Note that these protocols, in which the interaction strength decreases with time, are quite different from simple annealing procedures that correspond to slow cooling, or (equivalently) a gradual increase in interaction strength with time~\cite{rothemund06,ke12}.

\section{Outlook}
\label{sec:conc} 

We have shown how a very simple protocol for time-dependent interactions can significantly improve the self-assembly of crystals in a model colloidal system.  We have argued that this improvement is based on the physics of nucleation, growth and kinetic trapping.  Indeed, our results for a schematic model of cluster growth show a similar improvement in assembly yield when time-dependent interactions are used.  Physically, our central idea is that the best conditions for crystal growth are different from those for nucleation.  Such ideas have a long history and have been exploited in the protein crystallisation community (see Ref.~\onlinecite{vekilov99} and references therein).  However, we are not aware of simulation studies where the microscopic mechanisms of this effect are discussed in detail, nor of  applications of this principle in colloidal self-assembly.

For the successful application of this idea in self-assembly experiments, it is (obviously) essential that interparticle interactions can be manipulated in time, for example by a controllable depletion interaction~\cite{taylor12} or by temperature-dependent DNA-mediated interactions~\cite{nyk08,martinez11}.  Such experimental methods are in place, although fine control of the time-dependence may be challenging.  Certainly, the use of colloidal systems (where microscopic time scales may be of the order of milliseconds to seconds) brings with it different challenges to molecular systems where particles diffuse much more quickly.  In particular, the slower microscopic rates in colloidal systems can lead to greater propensity for kinetic trapping in disordered states. (When microscopic time scales are long, even relatively shallow traps can be relevant on experimental time scales of minutes to hours).  The use of time-dependent interactions may allow kinetic trapping to be controlled -- previous applications have focussed on the interplay between time-dependent interactinos and multiple nucleation events~\cite{vekilov99,schoen09}.

As well as their characteristic time scales, another feature of colloidal systems is that particles may be observed in real-time.  This offers the possibility for time-dependent interaction protocols that are selected on-the-fly, based on feedback from a system's behaviour.  Such ideas are beginning to be investigated~\cite{bevan12,klotsa-13,bevan16,truskett-arxiv16}, we look forward to further progress in this direction, which will require increased theoretical understanding of effects of time-dependent interactions, as well as creative new experimental ideas.

\begin{acknowledgments}
We thank Richard Sear and Steve Whitelam for helpful discussions, and the Engineering and Physical Sciences Research Council (EPSRC) for financial support through grant EP/L001438/1.
\end{acknowledgments}

\appendix

\section{Equilibrium and Quasiequilibrium in the schematic model}
\label{app:schem}

\newcommand{\cc}{\overline{c}}

This appendix collects some exact and approximate results for the schematic model of cluster growth that we introduced in Sec.~\ref{sec:schem-def}.

\subsection{Equilibrium average concentrations}
\label{app:equ}

From (\ref{equ:eq-const}) we can obtain the equilibrium concentrations of all species in terms of the free monomer concentration $\cc_0$, as
\begin{equation}
\cc({\rm g}_j) = \begin{cases} \cc_0 (x_\nu)^{j-1} , & j< \nc \\
                        \cc_0 (x_\nu)^{\nc-1} (x_1)^{j-\nc}, & j \geq \nc \end{cases}
\label{equ:ccg}
\end{equation}
and
\begin{equation}
\cc({\rm g}_{\nc+k} b_{k'}) = \cc_0 (x_\nu)^{\nc-1} (x_1)^k (x_\nu)^{k'}
\label{equ:ccb}
\end{equation}
with $x_\mu=(m\cc_0/\CR) \ee^{\eps/\mu}$, $x_\nu=(\cc_0/\CR) \ee^{\eps/\nu}$, and $x_1=(\cc_0/\CR) \ee^{\eps}$.

The sum-rule for particles (\ref{equ:sum-rule}) is then
\begin{equation}
\begin{split}
c_{\rm T} = & \cc_0 \sum_{k=1}^{\nc} k (x_\nu)^{k-1} \\ &+ \cc_0 \sum_{k=1}^\infty (\nc+k) (x_\nu)^{\nc-1} (x_1)^k
\\ & + 
\cc_0 \sum_{k=1}^\infty \sum_{k'=1}^\infty (\nc+k+k') (x_\mu)^{\nc-1} (x_1)^k (x_\mu)^{k'}
\end{split}
\label{equ:ct-all}
\end{equation}
where the first sum includes clusters of sizes $i< n_c$, the second includes clusters of sizes $i\geq n_c$ in which all particles are correctly bonded and the final sum includes all clusters containing incorrectly bonded particles.  Combining the last two sums, this equation may be written as
\begin{multline}
c_{\rm T} = \cc_0 \frac{\partial}{\partial \cc_0} \left[ \sum_{k=1}^{\nc} \cc_0(x_\nu)^{k-1} \right.
\\ 
\left. + 
\sum_{k=1}^\infty \sum_{k'=0}^\infty \cc_0(x_\nu)^{\nc-1} (x_1)^k (x_\mu)^{k'} \right]
\end{multline}
A solution is possible only if  $x_1,x_\nu<1$, in which case
the sums are straightforward geometrical series, and we obtain
\begin{multline}
c_{\rm T} = \cc_0 \frac{\partial}{\partial\cc_0} \left[ \frac{\cc_0(1-(x_\nu)^{\nc})}{1-x_\nu} 
+ 
 \frac{\cc_0 x_1 (x_\nu)^{\nc-1}}{(1-x_1)(1-x_\mu)} \right] .
 \label{equ:summed}
\end{multline}
After some algebra, this yields a polynomial of degree $n_c+1$ in $\cc_0$, which may be solved numerically.  Hence the concentrations of all species can be obtained from (\ref{equ:ccg},\ref{equ:ccb}).

Notice that the parameters $x_1$ and $x_\mu$ control the behaviours of the number of large correctly-bonded clusters and large incorrectly-bonded clusters, respectively.  For strong bonds, we expect $x_\mu<x_1$: in this case $x_1\to1^-$ corresponds to the system becoming dominated by large correctly-assembled clusters.  This is the analog of the thermodynamically-stable crystalline phase in the model colloid.  From the definition of $x_1$, one sees that $\cc_0\approx \CR \ee^{-\eps}$ in this limit.  Clearly if $c_{\rm T}<\CR \ee^{-\eps}$, the system is too dilute to achieve this limit (or the interactions are too weak): this provides an order-of-magnitude estimate for the onset of assembly, at $\eps^*\approx \ln (\CR/c_{\rm T})$, which is $\eps^* \approx 4.6$ for the parameters used in this work, consistent with Fig.~\ref{fig:yieldconste}(b).

\subsection{Equilibrium at finite $N$}

If the number of particles is large enough, the equilibrium concentration of any species in a typical configuration should be close to its expected (average) value.
However, in this work we consider the behaviour of this model for a fixed finite number of particles $N$.  This puts an upper limit on the size of all clusters, leading to cutoffs on the sums in (\ref{equ:ct-all}).  Moreoever, the number of clusters of species $X$ is $n(X)=Nc(X)/c_{\rm T}$ and the value of $n(X)$ in any configuration is (obviously) an integer, which restricts the possible values for $c(X)$.  In practice, when $\eps$ is large, a typical configuration in the equilibrium state is dominated by a single large cluster that contains a finite fraction of all the particles, with the remaining particles distributed mostly as free monomers. 

\subsection{Quasiequilibrium : pre-nucleation state}
\label{app:quasi-eq-pre-nuc}

As well as the stable equilibrium state, the model also includes a metastable (pre-nucleation) state.  Suppose that the sum rule (\ref{equ:ct-all}) is saturated by the first sum on the right hand side, so that the two infinite sums can be neglected.  This state is expected to be stable if the typical number of clusters of size $\nc$ is small compared to unity, since such intermediate-sized clusters are (obviously) required in order for the large clusters to grow.  This gives a self-consistency condition for metastability, $n({\rm g}_{\nc})=Nc({\rm g}_{\nc})/c_{\rm T}\ll 1$.  For an order-of-magnitude estimate, we assume that the metastable state is dominated by monomers: $c_0^{\rm qe}\approx c_{\rm T}$, where the label `qe' indicates quasiequilbrium.  In this case $c({\rm g}_{\nc}) \approx c_{\rm T} (c_{\rm T} \ee^{\eps/\nu}/\CR)^{\nc-1}$ so the condition for metastability is $\eps < \eps^{\rm qe}:=\nu [ \ln (\CR/c_{\rm T}) - (\ln N)/(n_c-1) ]$.  For the parameters considered here $\eps^{\rm qe}=13.8$. 

When the interaction strength $\eps>\eps^{\rm qe}$, the system quickly nucleates many clusters which then grow.  However, since there are so many clusters, the resulting concentration of free monomers is small, and subsequent cluster growth tends to be slow.  This last effect is particularly apparent in this model since there is no Ostwald ripening effect whereby larger clusters grow more quickly than small ones (there is no surface tension effect in the rate for unbinding from large clusters).  A closer agreement with the colloidal model could be obtained by including such an effect, via a more complex dependence of the rate $B_i$ on the cluster size $i$.  However, we do not consider this case here, for simplicity.

\subsection{Nucleation time}

Given the existence of a pre-nucleation state, it is natural to estimate the rate of nucleation.  In the case where nucleation is a rare event, this may be estimated as the probability of observing a cluster of size $\nc$, multiplied by the rate of growth of such clusters. (We assume that a cluster of size $\nc+1$ will grow quickly and not shrink, since the cluster has crossed the nucleation barrier.)  Following the argument of the previous section yields a rate $\tau_{\rm nuc}^{-1} \approx N F_1 c_{\rm T} (c_{\rm T}\ee^{\eps/\nu}/\CR)^{\nc-1}$, which is numerically small in the quasiequilibrium regime.

Alternatively one may follow Refs.~\onlinecite{zlotnick-2005,hagan-10} and consider an argument based on first-passage times for growing clusters.  Briefly, there are approximately $N$ clusters with sizes between $1$ and $\nc$, with growth rates $k_+=F_0 c_{\rm T}$ and shrinkage rates $k_-=F_0 \CR {\rm e}^{-\eps/\nu}$.  The size of each cluster follows a random walk with a reflecting boundary at size $0$.  We introduce an absorbing site at $\nc+1$: the idea is that the random walker gets absorbed when the cluster nucleates.  The mean time before absorbance for a random walk with rates $k_+,k_-$ and the requisite boundary conditions is (assuming $k_+\neq k_-$ and that the cluster starts at size $1$)~\cite{bar-haim-1998} 
\begin{equation}
\tau_{\rm abs} =
\frac{k_-}{(k_--k_+)^2}\left[ \left( \frac{k_-}{k_+} \right)^{\nc} - 1 \right] -\frac{\nc}{k_- - k_+} 
\end{equation}
Note that $k_- > k_+$ so this time is typically large.  Since there are $N$ clusters, the nucleation rate is $\tau_{\rm nuc}^{-1} \approx N/\tau_{\rm abs}$: in the limit where nucleation is rare we recover
\begin{equation}
\tau_{\rm nuc}^{-1} \approx \frac{ N k_+^{\nc} }{ k_-^{\nc-1} } = N F_1 c_{\rm T} \left( \frac{c_{\rm T}}{\CR} \ee^{\eps/\nu} \right)^{\nc-1} 
\end{equation}
consistent with the argument above based on the concentration of clusters of size $\nc$.

\subsection{Second quasiequilibrium state : kinetic trapping}

Since the schematic model was designed to account for both nucleation and kinetic trapping, it may be expected that kinetic trapping effects will be important after nucleation has taken place.  Recall that as soon as an incorrectly-bonded particle is added to a cluster, the cluster can only grow by incorrect bonding, unless all such particles are removed.  In practice this means that kinetic trapping dominates if the mean growth rate for incorrectly-bonded clusters $r_{\rm b}>0$, with
\begin{equation}r_{\rm b}=F_1[mc_0 - \CR\ee^{-\eps/\mu}].\end{equation}
The corresponding rate for correctly-bonded clusters is 
\begin{equation}r_{\rm g}=F_1[c_0 - \CR\ee^{-\eps}].\end{equation}

Just after nucleation, one expects $c_0\approx c_{\rm T}$ so $r_{\rm b}$ is positive if $\eps \gtrsim \eps^{\rm trap} = \mu \ln (\CR/mc_{\rm T})$.  For the parameters considered here $\eps^{\rm trap}=14.0$, very much comparable with $\eps^{\rm qe}$.  That is, kinetic trapping sets in for this system at around the same bond strength as nucleation ceases to be a rare event.  The maximum shown in Fig.~\ref{fig:yieldvare}(b) for the yield of the self-assembly process reflects the onset of kinetic trapping processes as well as the disappearance of the nucleation barrier (which tends to result in many small clusters instead of a small number of large ones).

As kinetically trapped clusters grow, the number of available monomers is reduced, and $r_{\rm b}$ decreases.
Eventually, the system reaches a quasi-equilibrium state with $r_{\rm b}\approx 0$.  In this state, $c_0\approx c_0^{\rm trap}:=(\CR/m) \ee^{-\eps/\mu}$. 
Note that in this regime $r_{\rm g}$ is still expected to be positive: the condition for (meta)stability of this quasi-equilibrium state is that there are no post-nucleation clusters to which correct binding is possible (all post-nucleation clusters include at least one incorrectly-bonded particle).  Escape from this metastable state typically takes place when a new correctly-bonded cluster nucleates: then one has $r_{\rm b}\approx 0$ and $r_{\rm g}>0$, so the new cluster tends to grow by correct binding.  This correctly-bonded cluster then further depletes the population $c_0$ of free monomers, leading to $r_{\rm b}<0$ and $r_{\rm g}>0$.  Hence the correctly bonded cluster tends to grow and take over the system: saturation happens when $r_{\rm g}\approx 0$ so that $c_0\approx\CR\ee^{-\eps}\approx c_0^{\rm eqm}$, and the system finally equilibrates.

\section{Method of Cluster Tracking}
\label{app:clusters}

A trajectory of the system consists of $M$ frames taken at times $(t_1,t_2,\dots,t_M)$.
To identify the clusters in each frame, we count particles that are mutually bonded together in isolated groups.
Between slices, many binding and unbinding events may have occured so, it is not trivial to identify which clusters at time $t_{j+1}$ are related to clusters at an earlier time $t_j$.
We use three criteria to identify causal connections between cluster $\alpha$ at time $t_{j}$ and cluster $\beta$ at time $t_{j+1}$.
The criteria are based on the principle that the clusters must contain a minimum number of the same particles (that is, there must be a shared `core' that surives all the binding and unbinding events).  Hence, to be causally connected:
\begin{enumerate}
	\item Both clusters must be larger than a cutoff size $N_{\rm min}$: that is, $N^{\alpha}_j,N^{\beta}_{j+1} > N_{\mathrm{min}}$.  We take $N_{\rm min}=52$.
	\item The number of particles that are shared by both clusters (the core) must represent a significant fraction of the smaller cluster: $N_{\rm core}/\mathrm{min}[N^{\alpha}_{j},N^{\beta}_{j+1}] > n_{\rm cut}$.  We take $n_{\rm cut}=\frac13$.
	\item The disparity between $N^{\alpha}_{j}$ and $N^{\beta}_{j+1}$ must not be too great: $\mathrm{max}[N^{\alpha}_{j}, N^{\beta}_{j+1}] < s\,\mathrm{min}[N^{\alpha}_{j}, N^{\beta}_{j+1}]$. We take $s = 4$.
\end{enumerate}

The first condition focusses attention on relatively large clusters, since these are the most important for understanding the dynamical evolution of the system.  The second condition accounts for the fact that two clusters may share a particle which evaporates from the surface of one cluster, spends some time as a free monomer, and subsequently binds to the second cluster.  In this case the two clusters are not causally connected.  Hence we require  that the clusters contain a significant number of shared particles.  The third condition means that when a small fragment breaks off from a large cluster, this is not interpreted as a causal connection: this criterion is useful since such events can be common but the resulting small clusters do not typically assemble further and do not contribute to the final product.  As a result, ignoring such links makes diagrams such as Fig.~\ref{fig:treeconste} less cluttered and easier to interpret.

In order to identify causal connections between trajectory frames that are not adjacent (for example betwen times $t_j$ and $t_{j+2}$, it is possible to form a rectangular matrix $A^j$ where the number of rows is the number of clusters at time $t_j$ and the number of columns is the number of clusters at time $t_{j+1}$.  The matrix element $(A^j)_{\alpha\beta}$ is equal to unity if cluster $\alpha$ at time $t_j$ is causally connected to cluster $\beta$ at time $t_{j+1}$.  Otherwise $(A^j)_{\alpha\beta}=0$.  In this case matrix multiplication allows identification of connections between times $t_j$ and $t_{j+2}$: two clusters $\alpha,\beta$ at these respective times are causally connected if $\sum_\gamma (A^j)_{\alpha\gamma} (A^{j+1})_{\gamma\beta}>0$.

This methodology allows the causal connections between clusters to be identified in Figs.~\ref{fig:treeconste},\ref{fig:treevare}.

\bibliography{var_str_int}

\begin{thebibliography}{54}%
\makeatletter
\providecommand \@ifxundefined [1]{%
 \@ifx{#1\undefined}
}%
\providecommand \@ifnum [1]{%
 \ifnum #1\expandafter \@firstoftwo
 \else \expandafter \@secondoftwo
 \fi
}%
\providecommand \@ifx [1]{%
 \ifx #1\expandafter \@firstoftwo
 \else \expandafter \@secondoftwo
 \fi
}%
\providecommand \natexlab [1]{#1}%
\providecommand \enquote  [1]{``#1''}%
\providecommand \bibnamefont  [1]{#1}%
\providecommand \bibfnamefont [1]{#1}%
\providecommand \citenamefont [1]{#1}%
\providecommand \href@noop [0]{\@secondoftwo}%
\providecommand \href [0]{\begingroup \@sanitize@url \@href}%
\providecommand \@href[1]{\@@startlink{#1}\@@href}%
\providecommand \@@href[1]{\endgroup#1\@@endlink}%
\providecommand \@sanitize@url [0]{\catcode `\\12\catcode `\$12\catcode
  `\&12\catcode `\#12\catcode `\^12\catcode `\_12\catcode `\%12\relax}%
\providecommand \@@startlink[1]{}%
\providecommand \@@endlink[0]{}%
\providecommand \url  [0]{\begingroup\@sanitize@url \@url }%
\providecommand \@url [1]{\endgroup\@href {#1}{\urlprefix }}%
\providecommand \urlprefix  [0]{URL }%
\providecommand \Eprint [0]{\href }%
\providecommand \doibase [0]{http://dx.doi.org/}%
\providecommand \selectlanguage [0]{\@gobble}%
\providecommand \bibinfo  [0]{\@secondoftwo}%
\providecommand \bibfield  [0]{\@secondoftwo}%
\providecommand \translation [1]{[#1]}%
\providecommand \BibitemOpen [0]{}%
\providecommand \bibitemStop [0]{}%
\providecommand \bibitemNoStop [0]{.\EOS\space}%
\providecommand \EOS [0]{\spacefactor3000\relax}%
\providecommand \BibitemShut  [1]{\csname bibitem#1\endcsname}%
\let\auto@bib@innerbib\@empty
\bibitem [{\citenamefont {Whitesides}\ and\ \citenamefont
  {Grzybowski}(2002)}]{white02review}%
  \BibitemOpen
  \bibfield  {author} {\bibinfo {author} {\bibfnamefont {G.~M.}\ \bibnamefont
  {Whitesides}}\ and\ \bibinfo {author} {\bibfnamefont {B.}~\bibnamefont
  {Grzybowski}},\ }\href@noop {} {\bibfield  {journal} {\bibinfo  {journal}
  {Science}\ }\textbf {\bibinfo {volume} {295}},\ \bibinfo {pages} {2418}
  (\bibinfo {year} {2002})}\BibitemShut {NoStop}%
\bibitem [{\citenamefont {Whitelam}\ and\ \citenamefont
  {Jack}(2015)}]{whitelam15-rev}%
  \BibitemOpen
  \bibfield  {author} {\bibinfo {author} {\bibfnamefont {S.}~\bibnamefont
  {Whitelam}}\ and\ \bibinfo {author} {\bibfnamefont {R.~L.}\ \bibnamefont
  {Jack}},\ }\href@noop {} {\bibfield  {journal} {\bibinfo  {journal} {Ann.
  Rev. Phys. Chem.}\ }\textbf {\bibinfo {volume} {66}},\ \bibinfo {pages} {142}
  (\bibinfo {year} {2015})}\BibitemShut {NoStop}%
\bibitem [{\citenamefont {Leunissen}\ \emph {et~al.}(2005)\citenamefont
  {Leunissen}, \citenamefont {Christova}, \citenamefont {Hynninen},
  \citenamefont {Royall}, \citenamefont {Campbell}, \citenamefont {Imhof},
  \citenamefont {Dijkstra}, \citenamefont {van Roij},\ and\ \citenamefont {van
  Blaaderen}}]{leunissen05}%
  \BibitemOpen
  \bibfield  {author} {\bibinfo {author} {\bibfnamefont {M.~E.}\ \bibnamefont
  {Leunissen}}, \bibinfo {author} {\bibfnamefont {C.~G.}\ \bibnamefont
  {Christova}}, \bibinfo {author} {\bibfnamefont {A.~P.}\ \bibnamefont
  {Hynninen}}, \bibinfo {author} {\bibfnamefont {C.~P.}\ \bibnamefont
  {Royall}}, \bibinfo {author} {\bibfnamefont {A.~I.}\ \bibnamefont
  {Campbell}}, \bibinfo {author} {\bibfnamefont {A.}~\bibnamefont {Imhof}},
  \bibinfo {author} {\bibfnamefont {M.}~\bibnamefont {Dijkstra}}, \bibinfo
  {author} {\bibfnamefont {R.}~\bibnamefont {van Roij}}, \ and\ \bibinfo
  {author} {\bibfnamefont {A.}~\bibnamefont {van Blaaderen}},\ }\href@noop {}
  {\bibfield  {journal} {\bibinfo  {journal} {Nature}\ }\textbf {\bibinfo
  {volume} {437}},\ \bibinfo {pages} {235} (\bibinfo {year}
  {2005})}\BibitemShut {NoStop}%
\bibitem [{\citenamefont {Hynninen}\ \emph {et~al.}(2007)\citenamefont
  {Hynninen}, \citenamefont {Thijssen}, \citenamefont {Vermolen}, \citenamefont
  {Dijkstra},\ and\ \citenamefont {Blaaderen}}]{Hynn2007}%
  \BibitemOpen
  \bibfield  {author} {\bibinfo {author} {\bibfnamefont {A.~P.}\ \bibnamefont
  {Hynninen}}, \bibinfo {author} {\bibfnamefont {J.~H.~J.}\ \bibnamefont
  {Thijssen}}, \bibinfo {author} {\bibfnamefont {E.~C.~M.}\ \bibnamefont
  {Vermolen}}, \bibinfo {author} {\bibfnamefont {M.}~\bibnamefont {Dijkstra}},
  \ and\ \bibinfo {author} {\bibfnamefont {A.~V.}\ \bibnamefont {Blaaderen}},\
  }\href@noop {} {\bibfield  {journal} {\bibinfo  {journal} {Nature Mat.}\
  }\textbf {\bibinfo {volume} {6}},\ \bibinfo {pages} {202} (\bibinfo {year}
  {2007})}\BibitemShut {NoStop}%
\bibitem [{\citenamefont {Sear}(2007)}]{sear07}%
  \BibitemOpen
  \bibfield  {author} {\bibinfo {author} {\bibfnamefont {R.~P.}\ \bibnamefont
  {Sear}},\ }\href@noop {} {\bibfield  {journal} {\bibinfo  {journal} {J.
  Phys.: Condens. Matter}\ }\textbf {\bibinfo {volume} {19}},\ \bibinfo {pages}
  {033101} (\bibinfo {year} {2007})}\BibitemShut {NoStop}%
\bibitem [{\citenamefont {Chen}, \citenamefont {Bae},\ and\ \citenamefont
  {Granick}(2011)}]{chen11}%
  \BibitemOpen
  \bibfield  {author} {\bibinfo {author} {\bibfnamefont {Q.}~\bibnamefont
  {Chen}}, \bibinfo {author} {\bibfnamefont {S.~C.}\ \bibnamefont {Bae}}, \
  and\ \bibinfo {author} {\bibfnamefont {S.}~\bibnamefont {Granick}},\
  }\href@noop {} {\bibfield  {journal} {\bibinfo  {journal} {Nature}\ }\textbf
  {\bibinfo {volume} {469}},\ \bibinfo {pages} {381} (\bibinfo {year}
  {2011})}\BibitemShut {NoStop}%
\bibitem [{\citenamefont {Romano}\ and\ \citenamefont
  {Sciortino}(2011)}]{romano11}%
  \BibitemOpen
  \bibfield  {author} {\bibinfo {author} {\bibfnamefont {F.}~\bibnamefont
  {Romano}}\ and\ \bibinfo {author} {\bibfnamefont {F.}~\bibnamefont
  {Sciortino}},\ }\href@noop {} {\bibfield  {journal} {\bibinfo  {journal}
  {Soft Matter}\ }\textbf {\bibinfo {volume} {7}},\ \bibinfo {pages} {5799}
  (\bibinfo {year} {2011})}\BibitemShut {NoStop}%
\bibitem [{\citenamefont {Klotsa}\ and\ \citenamefont
  {Jack}(2011)}]{klotsa-11}%
  \BibitemOpen
  \bibfield  {author} {\bibinfo {author} {\bibfnamefont {D.}~\bibnamefont
  {Klotsa}}\ and\ \bibinfo {author} {\bibfnamefont {R.~L.}\ \bibnamefont
  {Jack}},\ }\href@noop {} {\bibfield  {journal} {\bibinfo  {journal} {Soft
  Matter}\ }\textbf {\bibinfo {volume} {7}},\ \bibinfo {pages} {6294} (\bibinfo
  {year} {2011})}\BibitemShut {NoStop}%
\bibitem [{\citenamefont {Zlotnick}(2005)}]{zlotnick-2005}%
  \BibitemOpen
  \bibfield  {author} {\bibinfo {author} {\bibfnamefont {A.}~\bibnamefont
  {Zlotnick}},\ }\href@noop {} {\bibfield  {journal} {\bibinfo  {journal} {J.
  Mol. Rec.}\ }\textbf {\bibinfo {volume} {18}},\ \bibinfo {pages} {479}
  (\bibinfo {year} {2005})}\BibitemShut {NoStop}%
\bibitem [{\citenamefont {Perlmutter}\ and\ \citenamefont
  {Hagan}(2015)}]{hagan-annrev}%
  \BibitemOpen
  \bibfield  {author} {\bibinfo {author} {\bibfnamefont {J.~D.}\ \bibnamefont
  {Perlmutter}}\ and\ \bibinfo {author} {\bibfnamefont {M.~F.}\ \bibnamefont
  {Hagan}},\ }\href@noop {} {\bibfield  {journal} {\bibinfo  {journal} {Ann.
  Rev. Phys. Chem.}\ }\textbf {\bibinfo {volume} {66}},\ \bibinfo {pages} {217}
  (\bibinfo {year} {2015})}\BibitemShut {NoStop}%
\bibitem [{\citenamefont {Mirkin}\ \emph {et~al.}(1996)\citenamefont {Mirkin},
  \citenamefont {Letsinger}, \citenamefont {Mucic},\ and\ \citenamefont
  {Storhoff}}]{mirkin96}%
  \BibitemOpen
  \bibfield  {author} {\bibinfo {author} {\bibfnamefont {C.~A.}\ \bibnamefont
  {Mirkin}}, \bibinfo {author} {\bibfnamefont {R.~L.}\ \bibnamefont
  {Letsinger}}, \bibinfo {author} {\bibfnamefont {R.~C.}\ \bibnamefont
  {Mucic}}, \ and\ \bibinfo {author} {\bibfnamefont {J.~J.}\ \bibnamefont
  {Storhoff}},\ }\href@noop {} {\bibfield  {journal} {\bibinfo  {journal}
  {Nature}\ }\textbf {\bibinfo {volume} {382}},\ \bibinfo {pages} {607}
  (\bibinfo {year} {1996})}\BibitemShut {NoStop}%
\bibitem [{\citenamefont {Rothemund}(2006)}]{rothemund06}%
  \BibitemOpen
  \bibfield  {author} {\bibinfo {author} {\bibfnamefont {P.~W.~K.}\
  \bibnamefont {Rothemund}},\ }\href@noop {} {\bibfield  {journal} {\bibinfo
  {journal} {Nature}\ }\textbf {\bibinfo {volume} {440}},\ \bibinfo {pages}
  {297} (\bibinfo {year} {2006})}\BibitemShut {NoStop}%
\bibitem [{\citenamefont {Macfarlane}\ \emph {et~al.}(2011)\citenamefont
  {Macfarlane}, \citenamefont {Lee}, \citenamefont {Jones}, \citenamefont
  {Harris}, \citenamefont {Schartz},\ and\ \citenamefont
  {Mirkin}}]{macfarlane11}%
  \BibitemOpen
  \bibfield  {author} {\bibinfo {author} {\bibfnamefont {R.~J.}\ \bibnamefont
  {Macfarlane}}, \bibinfo {author} {\bibfnamefont {B.}~\bibnamefont {Lee}},
  \bibinfo {author} {\bibfnamefont {M.~R.}\ \bibnamefont {Jones}}, \bibinfo
  {author} {\bibfnamefont {N.}~\bibnamefont {Harris}}, \bibinfo {author}
  {\bibfnamefont {G.~C.}\ \bibnamefont {Schartz}}, \ and\ \bibinfo {author}
  {\bibfnamefont {C.~A.}\ \bibnamefont {Mirkin}},\ }\href@noop {} {\bibfield
  {journal} {\bibinfo  {journal} {Science}\ }\textbf {\bibinfo {volume}
  {334}},\ \bibinfo {pages} {204} (\bibinfo {year} {2011})}\BibitemShut
  {NoStop}%
\bibitem [{\citenamefont {Nykypanchuk}\ \emph {et~al.}(2008)\citenamefont
  {Nykypanchuk}, \citenamefont {Maye}, \citenamefont {van~der Lelie},\ and\
  \citenamefont {Gang}}]{nyk08}%
  \BibitemOpen
  \bibfield  {author} {\bibinfo {author} {\bibfnamefont {O.}~\bibnamefont
  {Nykypanchuk}}, \bibinfo {author} {\bibfnamefont {M.}~\bibnamefont {Maye}},
  \bibinfo {author} {\bibfnamefont {D.}~\bibnamefont {van~der Lelie}}, \ and\
  \bibinfo {author} {\bibfnamefont {O.}~\bibnamefont {Gang}},\ }\href@noop {}
  {\bibfield  {journal} {\bibinfo  {journal} {Nature}\ }\textbf {\bibinfo
  {volume} {451}},\ \bibinfo {pages} {549} (\bibinfo {year}
  {2008})}\BibitemShut {NoStop}%
\bibitem [{\citenamefont {Ke}\ \emph {et~al.}(2012)\citenamefont {Ke},
  \citenamefont {Ong}, \citenamefont {Shih},\ and\ \citenamefont {Yin}}]{ke12}%
  \BibitemOpen
  \bibfield  {author} {\bibinfo {author} {\bibfnamefont {Y.}~\bibnamefont
  {Ke}}, \bibinfo {author} {\bibfnamefont {L.~L.}\ \bibnamefont {Ong}},
  \bibinfo {author} {\bibfnamefont {W.~M.}\ \bibnamefont {Shih}}, \ and\
  \bibinfo {author} {\bibfnamefont {P.}~\bibnamefont {Yin}},\ }\href@noop {}
  {\bibfield  {journal} {\bibinfo  {journal} {Science}\ }\textbf {\bibinfo
  {volume} {338}},\ \bibinfo {pages} {1177} (\bibinfo {year}
  {2012})}\BibitemShut {NoStop}%
\bibitem [{\citenamefont {Glotzer}\ and\ \citenamefont
  {Solomon}(2007)}]{sol07}%
  \BibitemOpen
  \bibfield  {author} {\bibinfo {author} {\bibfnamefont {S.~C.}\ \bibnamefont
  {Glotzer}}\ and\ \bibinfo {author} {\bibfnamefont {M.~J.}\ \bibnamefont
  {Solomon}},\ }\href@noop {} {\bibfield  {journal} {\bibinfo  {journal}
  {Nature Mat.}\ }\textbf {\bibinfo {volume} {6}},\ \bibinfo {pages} {557}
  (\bibinfo {year} {2007})}\BibitemShut {NoStop}%
\bibitem [{\citenamefont {Pawar}\ and\ \citenamefont
  {Kretzschmar}(2009)}]{pawar09}%
  \BibitemOpen
  \bibfield  {author} {\bibinfo {author} {\bibfnamefont {A.~B.}\ \bibnamefont
  {Pawar}}\ and\ \bibinfo {author} {\bibfnamefont {I.}~\bibnamefont
  {Kretzschmar}},\ }\href@noop {} {\bibfield  {journal} {\bibinfo  {journal}
  {Langmuir}\ }\textbf {\bibinfo {volume} {25}},\ \bibinfo {pages} {9057}
  (\bibinfo {year} {2009})}\BibitemShut {NoStop}%
\bibitem [{\citenamefont {Sacanna}\ \emph {et~al.}(2010)\citenamefont
  {Sacanna}, \citenamefont {Irvine}, \citenamefont {Chaikin},\ and\
  \citenamefont {Pine}}]{sac10}%
  \BibitemOpen
  \bibfield  {author} {\bibinfo {author} {\bibfnamefont {S.}~\bibnamefont
  {Sacanna}}, \bibinfo {author} {\bibfnamefont {W.~T.~M.}\ \bibnamefont
  {Irvine}}, \bibinfo {author} {\bibfnamefont {P.~M.}\ \bibnamefont {Chaikin}},
  \ and\ \bibinfo {author} {\bibfnamefont {D.~J.}\ \bibnamefont {Pine}},\
  }\href@noop {} {\bibfield  {journal} {\bibinfo  {journal} {Nature}\ }\textbf
  {\bibinfo {volume} {464}},\ \bibinfo {pages} {575} (\bibinfo {year}
  {2010})}\BibitemShut {NoStop}%
\bibitem [{\citenamefont {Jiang}\ \emph {et~al.}(2010)\citenamefont {Jiang},
  \citenamefont {Chen}, \citenamefont {Tripathy}, \citenamefont {Luijten},
  \citenamefont {Schwiezer},\ and\ \citenamefont {Granick}}]{jiang10}%
  \BibitemOpen
  \bibfield  {author} {\bibinfo {author} {\bibfnamefont {S.}~\bibnamefont
  {Jiang}}, \bibinfo {author} {\bibfnamefont {Q.}~\bibnamefont {Chen}},
  \bibinfo {author} {\bibfnamefont {M.}~\bibnamefont {Tripathy}}, \bibinfo
  {author} {\bibfnamefont {E.}~\bibnamefont {Luijten}}, \bibinfo {author}
  {\bibfnamefont {K.~S.}\ \bibnamefont {Schwiezer}}, \ and\ \bibinfo {author}
  {\bibfnamefont {S.}~\bibnamefont {Granick}},\ }\href@noop {} {\bibfield
  {journal} {\bibinfo  {journal} {Advanced Materials}\ }\textbf {\bibinfo
  {volume} {22}},\ \bibinfo {pages} {1060} (\bibinfo {year}
  {2010})}\BibitemShut {NoStop}%
\bibitem [{\citenamefont {Kraft}\ \emph {et~al.}(2012)\citenamefont {Kraft},
  \citenamefont {Ni}, \citenamefont {Smallenburg}, \citenamefont {Hermes},
  \citenamefont {Yoon}, \citenamefont {Weitz}, \citenamefont {Blaaderen},
  \citenamefont {Groenewold}, \citenamefont {Dijkstra},\ and\ \citenamefont
  {Kegel}}]{kraft12}%
  \BibitemOpen
  \bibfield  {author} {\bibinfo {author} {\bibfnamefont {D.~J.}\ \bibnamefont
  {Kraft}}, \bibinfo {author} {\bibfnamefont {R.}~\bibnamefont {Ni}}, \bibinfo
  {author} {\bibfnamefont {F.}~\bibnamefont {Smallenburg}}, \bibinfo {author}
  {\bibfnamefont {M.}~\bibnamefont {Hermes}}, \bibinfo {author} {\bibfnamefont
  {K.}~\bibnamefont {Yoon}}, \bibinfo {author} {\bibfnamefont {D.~A.}\
  \bibnamefont {Weitz}}, \bibinfo {author} {\bibfnamefont {A.~V.}\ \bibnamefont
  {Blaaderen}}, \bibinfo {author} {\bibfnamefont {J.}~\bibnamefont
  {Groenewold}}, \bibinfo {author} {\bibfnamefont {M.}~\bibnamefont
  {Dijkstra}}, \ and\ \bibinfo {author} {\bibfnamefont {W.}~\bibnamefont
  {Kegel}},\ }\href@noop {} {\bibfield  {journal} {\bibinfo  {journal} {PNAS}\
  }\textbf {\bibinfo {volume} {109}},\ \bibinfo {pages} {10787} (\bibinfo
  {year} {2012})}\BibitemShut {NoStop}%
\bibitem [{\citenamefont {Sacanna}\ \emph {et~al.}(2013)\citenamefont
  {Sacanna}, \citenamefont {Korpics}, \citenamefont {Rodriguez}, \citenamefont
  {Col\'on-Mel\'endez}, \citenamefont {Kim}, \citenamefont {Pine},\ and\
  \citenamefont {Gi-Ra}}]{sacanna13}%
  \BibitemOpen
  \bibfield  {author} {\bibinfo {author} {\bibfnamefont {S.}~\bibnamefont
  {Sacanna}}, \bibinfo {author} {\bibfnamefont {M.}~\bibnamefont {Korpics}},
  \bibinfo {author} {\bibfnamefont {K.}~\bibnamefont {Rodriguez}}, \bibinfo
  {author} {\bibfnamefont {L.}~\bibnamefont {Col\'on-Mel\'endez}}, \bibinfo
  {author} {\bibfnamefont {S.~H.}\ \bibnamefont {Kim}}, \bibinfo {author}
  {\bibfnamefont {D.~J.}\ \bibnamefont {Pine}}, \ and\ \bibinfo {author}
  {\bibfnamefont {Y.}~\bibnamefont {Gi-Ra}},\ }\href@noop {} {\bibfield
  {journal} {\bibinfo  {journal} {Nature Communications}\ }\textbf {\bibinfo
  {volume} {4}},\ \bibinfo {pages} {1688} (\bibinfo {year} {2013})}\BibitemShut
  {NoStop}%
\bibitem [{\citenamefont {Wilber}, \citenamefont {Doye},\ and\ \citenamefont
  {Louis}(2009)}]{wilber09}%
  \BibitemOpen
  \bibfield  {author} {\bibinfo {author} {\bibfnamefont {A.~W.}\ \bibnamefont
  {Wilber}}, \bibinfo {author} {\bibfnamefont {J.~P.~K.}\ \bibnamefont {Doye}},
  \ and\ \bibinfo {author} {\bibfnamefont {A.~A.}\ \bibnamefont {Louis}},\
  }\href@noop {} {\bibfield  {journal} {\bibinfo  {journal} {J. Chem. Phys.}\
  }\textbf {\bibinfo {volume} {131}},\ \bibinfo {pages} {175101} (\bibinfo
  {year} {2009})}\BibitemShut {NoStop}%
\bibitem [{\citenamefont {Haji-Akbari}\ \emph {et~al.}(2009)\citenamefont
  {Haji-Akbari}, \citenamefont {Engel}, \citenamefont {Keys}, \citenamefont
  {Zheng}, \citenamefont {Petschek}, \citenamefont {Palffy-Muhoray},\ and\
  \citenamefont {Glotzer}}]{akbari09}%
  \BibitemOpen
  \bibfield  {author} {\bibinfo {author} {\bibfnamefont {A.}~\bibnamefont
  {Haji-Akbari}}, \bibinfo {author} {\bibfnamefont {M.}~\bibnamefont {Engel}},
  \bibinfo {author} {\bibfnamefont {A.~S.}\ \bibnamefont {Keys}}, \bibinfo
  {author} {\bibfnamefont {X.}~\bibnamefont {Zheng}}, \bibinfo {author}
  {\bibfnamefont {R.~G.}\ \bibnamefont {Petschek}}, \bibinfo {author}
  {\bibfnamefont {P.}~\bibnamefont {Palffy-Muhoray}}, \ and\ \bibinfo {author}
  {\bibfnamefont {S.~C.}\ \bibnamefont {Glotzer}},\ }\href@noop {} {\bibfield
  {journal} {\bibinfo  {journal} {Nature}\ }\textbf {\bibinfo {volume} {462}},\
  \bibinfo {pages} {773} (\bibinfo {year} {2009})}\BibitemShut {NoStop}%
\bibitem [{\citenamefont {Martinez-Veracoechea}\ \emph
  {et~al.}(2011)\citenamefont {Martinez-Veracoechea}, \citenamefont {Mladek},
  \citenamefont {Tkachenko},\ and\ \citenamefont {Frenkel}}]{martinez11}%
  \BibitemOpen
  \bibfield  {author} {\bibinfo {author} {\bibfnamefont {J.~F.}\ \bibnamefont
  {Martinez-Veracoechea}}, \bibinfo {author} {\bibfnamefont {B.~M.}\
  \bibnamefont {Mladek}}, \bibinfo {author} {\bibfnamefont {A.~V.}\
  \bibnamefont {Tkachenko}}, \ and\ \bibinfo {author} {\bibfnamefont
  {D.}~\bibnamefont {Frenkel}},\ }\href@noop {} {\bibfield  {journal} {\bibinfo
   {journal} {Phys. Rev. Lett.}\ }\textbf {\bibinfo {volume} {107}},\ \bibinfo
  {pages} {04592} (\bibinfo {year} {2011})}\BibitemShut {NoStop}%
\bibitem [{\citenamefont {Hagan}\ and\ \citenamefont
  {Chandler}(2006)}]{hagan06}%
  \BibitemOpen
  \bibfield  {author} {\bibinfo {author} {\bibfnamefont {M.~F.}\ \bibnamefont
  {Hagan}}\ and\ \bibinfo {author} {\bibfnamefont {D.}~\bibnamefont
  {Chandler}},\ }\href@noop {} {\bibfield  {journal} {\bibinfo  {journal}
  {Biophys. J.}\ }\textbf {\bibinfo {volume} {91}},\ \bibinfo {pages} {42}
  (\bibinfo {year} {2006})}\BibitemShut {NoStop}%
\bibitem [{\citenamefont {Rapaport}(2008)}]{rap08}%
  \BibitemOpen
  \bibfield  {author} {\bibinfo {author} {\bibfnamefont {D.~C.}\ \bibnamefont
  {Rapaport}},\ }\href@noop {} {\bibfield  {journal} {\bibinfo  {journal}
  {Phys. Rev. Lett.}\ }\textbf {\bibinfo {volume} {101}},\ \bibinfo {pages}
  {186101} (\bibinfo {year} {2008})}\BibitemShut {NoStop}%
\bibitem [{\citenamefont {Whitesides}\ and\ \citenamefont
  {Boncheva}(2002)}]{white02}%
  \BibitemOpen
  \bibfield  {author} {\bibinfo {author} {\bibfnamefont {G.~M.}\ \bibnamefont
  {Whitesides}}\ and\ \bibinfo {author} {\bibfnamefont {M.}~\bibnamefont
  {Boncheva}},\ }\href@noop {} {\bibfield  {journal} {\bibinfo  {journal}
  {PNAS}\ }\textbf {\bibinfo {volume} {99}},\ \bibinfo {pages} {4769} (\bibinfo
  {year} {2002})}\BibitemShut {NoStop}%
\bibitem [{\citenamefont {Whitelam}\ \emph {et~al.}(2009)\citenamefont
  {Whitelam}, \citenamefont {Feng}, \citenamefont {Hagan},\ and\ \citenamefont
  {Geissler}}]{whitelam09}%
  \BibitemOpen
  \bibfield  {author} {\bibinfo {author} {\bibfnamefont {S.}~\bibnamefont
  {Whitelam}}, \bibinfo {author} {\bibfnamefont {E.~H.}\ \bibnamefont {Feng}},
  \bibinfo {author} {\bibfnamefont {M.~F.}\ \bibnamefont {Hagan}}, \ and\
  \bibinfo {author} {\bibfnamefont {P.~L.}\ \bibnamefont {Geissler}},\
  }\href@noop {} {\bibfield  {journal} {\bibinfo  {journal} {Soft Matter}\
  }\textbf {\bibinfo {volume} {5}},\ \bibinfo {pages} {1251} (\bibinfo {year}
  {2009})}\BibitemShut {NoStop}%
\bibitem [{\citenamefont {Grant}, \citenamefont {Jack},\ and\ \citenamefont
  {Whitelam}(2011)}]{grant-11}%
  \BibitemOpen
  \bibfield  {author} {\bibinfo {author} {\bibfnamefont {J.}~\bibnamefont
  {Grant}}, \bibinfo {author} {\bibfnamefont {R.~L.}\ \bibnamefont {Jack}}, \
  and\ \bibinfo {author} {\bibfnamefont {S.}~\bibnamefont {Whitelam}},\
  }\href@noop {} {\bibfield  {journal} {\bibinfo  {journal} {J. Chem. Phys.}\
  }\textbf {\bibinfo {volume} {135}},\ \bibinfo {pages} {214505} (\bibinfo
  {year} {2011})}\BibitemShut {NoStop}%
\bibitem [{\citenamefont {Galkin}\ and\ \citenamefont
  {Vekilov}(1999)}]{vekilov99}%
  \BibitemOpen
  \bibfield  {author} {\bibinfo {author} {\bibfnamefont {O.}~\bibnamefont
  {Galkin}}\ and\ \bibinfo {author} {\bibfnamefont {P.~G.}\ \bibnamefont
  {Vekilov}},\ }\href@noop {} {\bibfield  {journal} {\bibinfo  {journal} {J.
  Phys. Chem. B}\ }\textbf {\bibinfo {volume} {103}},\ \bibinfo {pages} {10965}
  (\bibinfo {year} {1999})}\BibitemShut {NoStop}%
\bibitem [{\citenamefont {Schoen}(2009)}]{schoen09}%
  \BibitemOpen
  \bibfield  {author} {\bibinfo {author} {\bibfnamefont {J.~C.}\ \bibnamefont
  {Schoen}},\ }\href@noop {} {\bibfield  {journal} {\bibinfo  {journal} {Z.
  Anorg. Allg. Chem.}\ }\textbf {\bibinfo {volume} {635}},\ \bibinfo {pages}
  {1794} (\bibinfo {year} {2009})}\BibitemShut {NoStop}%
\bibitem [{\citenamefont {Klotsa}\ and\ \citenamefont
  {Jack}(2013)}]{klotsa-13}%
  \BibitemOpen
  \bibfield  {author} {\bibinfo {author} {\bibfnamefont {D.}~\bibnamefont
  {Klotsa}}\ and\ \bibinfo {author} {\bibfnamefont {R.~L.}\ \bibnamefont
  {Jack}},\ }\href@noop {} {\bibfield  {journal} {\bibinfo  {journal} {J. Chem.
  Phys.}\ }\textbf {\bibinfo {volume} {138}},\ \bibinfo {pages} {094502}
  (\bibinfo {year} {2013})}\BibitemShut {NoStop}%
\bibitem [{\citenamefont {Taylor}, \citenamefont {Evans},\ and\ \citenamefont
  {Royal}(2012)}]{taylor12}%
  \BibitemOpen
  \bibfield  {author} {\bibinfo {author} {\bibfnamefont {S.~L.}\ \bibnamefont
  {Taylor}}, \bibinfo {author} {\bibfnamefont {R.}~\bibnamefont {Evans}}, \
  and\ \bibinfo {author} {\bibfnamefont {C.~P.}\ \bibnamefont {Royal}},\
  }\href@noop {} {\bibfield  {journal} {\bibinfo  {journal} {J. Phys.: Condens.
  Matter}\ }\textbf {\bibinfo {volume} {24}},\ \bibinfo {pages} {464128}
  (\bibinfo {year} {2012})}\BibitemShut {NoStop}%
\bibitem [{vmd()}]{vmd}%
  \BibitemOpen
  \href@noop {} {}\bibinfo {note} {Snapshots created using VMD,\\ {\tt
  http://www.ks.uiuc.edu/Research/vmd/}}\BibitemShut {NoStop}%
\bibitem [{\citenamefont {Poon}(2002)}]{Poon02}%
  \BibitemOpen
  \bibfield  {author} {\bibinfo {author} {\bibfnamefont {W.~C.~K.}\
  \bibnamefont {Poon}},\ }\href {http://stacks.iop.org/0953-8984/14/i=33/a=201}
  {\bibfield  {journal} {\bibinfo  {journal} {J. Phys.: Condens. Matter}\
  }\textbf {\bibinfo {volume} {14}},\ \bibinfo {pages} {R859} (\bibinfo {year}
  {2002})}\BibitemShut {NoStop}%
\bibitem [{\citenamefont {Haxton}, \citenamefont {Hedges},\ and\ \citenamefont
  {Whitelam}(2015)}]{haxton15}%
  \BibitemOpen
  \bibfield  {author} {\bibinfo {author} {\bibfnamefont {T.~K.}\ \bibnamefont
  {Haxton}}, \bibinfo {author} {\bibfnamefont {L.~O.}\ \bibnamefont {Hedges}},
  \ and\ \bibinfo {author} {\bibfnamefont {S.}~\bibnamefont {Whitelam}},\
  }\href@noop {} {\bibfield  {journal} {\bibinfo  {journal} {Soft Matter}\
  }\textbf {\bibinfo {volume} {11}},\ \bibinfo {pages} {9307} (\bibinfo {year}
  {2015})}\BibitemShut {NoStop}%
\bibitem [{\citenamefont {Noro}\ and\ \citenamefont
  {Frenkel}(2000)}]{noro2000}%
  \BibitemOpen
  \bibfield  {author} {\bibinfo {author} {\bibfnamefont {M.~G.}\ \bibnamefont
  {Noro}}\ and\ \bibinfo {author} {\bibfnamefont {D.}~\bibnamefont {Frenkel}},\
  }\href@noop {} {\bibfield  {journal} {\bibinfo  {journal} {J. Chem. Phys.}\
  }\textbf {\bibinfo {volume} {113}},\ \bibinfo {pages} {2941} (\bibinfo {year}
  {2000})}\BibitemShut {NoStop}%
\bibitem [{\citenamefont {Rosenbaum}, \citenamefont {Zamora},\ and\
  \citenamefont {Zukoski}(1996)}]{rosenbaum96}%
  \BibitemOpen
  \bibfield  {author} {\bibinfo {author} {\bibfnamefont {D.~F.}\ \bibnamefont
  {Rosenbaum}}, \bibinfo {author} {\bibfnamefont {P.~C.}\ \bibnamefont
  {Zamora}}, \ and\ \bibinfo {author} {\bibfnamefont {C.~F.}\ \bibnamefont
  {Zukoski}},\ }\href@noop {} {\bibfield  {journal} {\bibinfo  {journal} {Phys.
  Rev. Lett.}\ }\textbf {\bibinfo {volume} {76}},\ \bibinfo {pages} {150}
  (\bibinfo {year} {1996})}\BibitemShut {NoStop}%
\bibitem [{\citenamefont {Rosenbaum}\ \emph {et~al.}(1999)\citenamefont
  {Rosenbaum}, \citenamefont {Kulkarni}, \citenamefont {Ramakrishnan},\ and\
  \citenamefont {Zukoski}}]{rosenbaum99}%
  \BibitemOpen
  \bibfield  {author} {\bibinfo {author} {\bibfnamefont {D.~F.}\ \bibnamefont
  {Rosenbaum}}, \bibinfo {author} {\bibfnamefont {A.}~\bibnamefont {Kulkarni}},
  \bibinfo {author} {\bibfnamefont {S.}~\bibnamefont {Ramakrishnan}}, \ and\
  \bibinfo {author} {\bibfnamefont {C.~F.}\ \bibnamefont {Zukoski}},\
  }\href@noop {} {\bibfield  {journal} {\bibinfo  {journal} {J. Chem. Phys.}\
  }\textbf {\bibinfo {volume} {111}},\ \bibinfo {pages} {9882} (\bibinfo {year}
  {1999})}\BibitemShut {NoStop}%
\bibitem [{\citenamefont {Frenkel}\ and\ \citenamefont
  {Smit}(2001)}]{frenkelsmit}%
  \BibitemOpen
  \bibfield  {author} {\bibinfo {author} {\bibfnamefont {D.}~\bibnamefont
  {Frenkel}}\ and\ \bibinfo {author} {\bibfnamefont {B.}~\bibnamefont {Smit}},\
  }\href@noop {} {\emph {\bibinfo {title} {Understanding Molecular Simulation:
  From Algorithms to Applications}}}\ (\bibinfo  {publisher} {Elsevier},\
  \bibinfo {year} {2001})\BibitemShut {NoStop}%
\bibitem [{\citenamefont {Whitelam}(2011)}]{whitelam11-molsim}%
  \BibitemOpen
  \bibfield  {author} {\bibinfo {author} {\bibfnamefont {S.}~\bibnamefont
  {Whitelam}},\ }\href@noop {} {\bibfield  {journal} {\bibinfo  {journal} {Mol.
  Sim.}\ }\textbf {\bibinfo {volume} {37}},\ \bibinfo {pages} {606} (\bibinfo
  {year} {2011})}\BibitemShut {NoStop}%
\bibitem [{\citenamefont {Fortini}, \citenamefont {Sanz},\ and\ \citenamefont
  {Dijkstra}(2008)}]{fortini08}%
  \BibitemOpen
  \bibfield  {author} {\bibinfo {author} {\bibfnamefont {A.}~\bibnamefont
  {Fortini}}, \bibinfo {author} {\bibfnamefont {E.}~\bibnamefont {Sanz}}, \
  and\ \bibinfo {author} {\bibfnamefont {M.}~\bibnamefont {Dijkstra}},\
  }\href@noop {} {\bibfield  {journal} {\bibinfo  {journal} {Phys. Rev. E}\
  }\textbf {\bibinfo {volume} {78}},\ \bibinfo {pages} {041402} (\bibinfo
  {year} {2008})}\BibitemShut {NoStop}%
\bibitem [{\citenamefont {Becker}\ and\ \citenamefont
  {Doering}(1935)}]{becker35}%
  \BibitemOpen
  \bibfield  {author} {\bibinfo {author} {\bibfnamefont {R.}~\bibnamefont
  {Becker}}\ and\ \bibinfo {author} {\bibfnamefont {W.}~\bibnamefont
  {Doering}},\ }\href@noop {} {\bibfield  {journal} {\bibinfo  {journal} {Ann.
  Phys}\ }\textbf {\bibinfo {volume} {24}},\ \bibinfo {pages} {719} (\bibinfo
  {year} {1935})}\BibitemShut {NoStop}%
\bibitem [{\citenamefont {Binder}\ and\ \citenamefont
  {Stauffer}(1976)}]{binder76}%
  \BibitemOpen
  \bibfield  {author} {\bibinfo {author} {\bibfnamefont {K.}~\bibnamefont
  {Binder}}\ and\ \bibinfo {author} {\bibfnamefont {D.}~\bibnamefont
  {Stauffer}},\ }\href@noop {} {\bibfield  {journal} {\bibinfo  {journal} {Adv.
  Phys.}\ }\textbf {\bibinfo {volume} {25}},\ \bibinfo {pages} {343} (\bibinfo
  {year} {1976})}\BibitemShut {NoStop}%
\bibitem [{\citenamefont {Hagan}\ and\ \citenamefont {Elrad}(2010)}]{hagan-10}%
  \BibitemOpen
  \bibfield  {author} {\bibinfo {author} {\bibfnamefont {M.~F.}\ \bibnamefont
  {Hagan}}\ and\ \bibinfo {author} {\bibfnamefont {O.~M.}\ \bibnamefont
  {Elrad}},\ }\href@noop {} {\bibfield  {journal} {\bibinfo  {journal}
  {Biophys. J.}\ }\textbf {\bibinfo {volume} {98}},\ \bibinfo {pages} {1065}
  (\bibinfo {year} {2010})}\BibitemShut {NoStop}%
\bibitem [{\citenamefont {Knowles}\ \emph {et~al.}(2009)\citenamefont
  {Knowles}, \citenamefont {Waudby}, \citenamefont {Devlin}, \citenamefont
  {Cohen}, \citenamefont {Aguzzi}, \citenamefont {Vendruscolo}, \citenamefont
  {Terentjev}, \citenamefont {Welland},\ and\ \citenamefont
  {Dobson}}]{knowles2009}%
  \BibitemOpen
  \bibfield  {author} {\bibinfo {author} {\bibfnamefont {T.~P.~J.}\
  \bibnamefont {Knowles}}, \bibinfo {author} {\bibfnamefont {C.~A.}\
  \bibnamefont {Waudby}}, \bibinfo {author} {\bibfnamefont {G.~L.}\
  \bibnamefont {Devlin}}, \bibinfo {author} {\bibfnamefont {S.~I.~A.}\
  \bibnamefont {Cohen}}, \bibinfo {author} {\bibfnamefont {A.}~\bibnamefont
  {Aguzzi}}, \bibinfo {author} {\bibfnamefont {M.}~\bibnamefont {Vendruscolo}},
  \bibinfo {author} {\bibfnamefont {E.~M.}\ \bibnamefont {Terentjev}}, \bibinfo
  {author} {\bibfnamefont {M.~E.}\ \bibnamefont {Welland}}, \ and\ \bibinfo
  {author} {\bibfnamefont {C.~M.}\ \bibnamefont {Dobson}},\ }\href@noop {}
  {\bibfield  {journal} {\bibinfo  {journal} {Science}\ }\textbf {\bibinfo
  {volume} {326}},\ \bibinfo {pages} {1533} (\bibinfo {year}
  {2009})}\BibitemShut {NoStop}%
\bibitem [{\citenamefont {Whitelam}, \citenamefont {Dahal},\ and\ \citenamefont
  {Schmit}(2016)}]{whitelam16-poison}%
  \BibitemOpen
  \bibfield  {author} {\bibinfo {author} {\bibfnamefont {S.}~\bibnamefont
  {Whitelam}}, \bibinfo {author} {\bibfnamefont {Y.~R.}\ \bibnamefont {Dahal}},
  \ and\ \bibinfo {author} {\bibfnamefont {J.~D.}\ \bibnamefont {Schmit}},\
  }\href@noop {} {\bibfield  {journal} {\bibinfo  {journal} {J. Chem. Phys.}\
  }\textbf {\bibinfo {volume} {144}},\ \bibinfo {pages} {064903} (\bibinfo
  {year} {2016})}\BibitemShut {NoStop}%
\bibitem [{\citenamefont {Mithen}\ and\ \citenamefont {Sear}(2016)}]{Mithen16}%
  \BibitemOpen
  \bibfield  {author} {\bibinfo {author} {\bibfnamefont {J.~P.}\ \bibnamefont
  {Mithen}}\ and\ \bibinfo {author} {\bibfnamefont {R.~P.}\ \bibnamefont
  {Sear}},\ }\href@noop {} {\bibfield  {journal} {\bibinfo  {journal} {Crystal
  Growth and Design}\ }\textbf {\bibinfo {volume} {16}},\ \bibinfo {pages}
  {3049} (\bibinfo {year} {2016})}\BibitemShut {NoStop}%
\bibitem [{\citenamefont {ten Wolde}, \citenamefont {Ruiz-Montero},\ and\
  \citenamefont {Frenkel}(1995)}]{tenwolde95}%
  \BibitemOpen
  \bibfield  {author} {\bibinfo {author} {\bibfnamefont {P.~R.}\ \bibnamefont
  {ten Wolde}}, \bibinfo {author} {\bibfnamefont {M.~J.}\ \bibnamefont
  {Ruiz-Montero}}, \ and\ \bibinfo {author} {\bibfnamefont {D.}~\bibnamefont
  {Frenkel}},\ }\href@noop {} {\bibfield  {journal} {\bibinfo  {journal} {Phys.
  Rev. Lett.}\ }\textbf {\bibinfo {volume} {75}},\ \bibinfo {pages} {2714}
  (\bibinfo {year} {1995})}\BibitemShut {NoStop}%
\bibitem [{\citenamefont {Jankowski}\ and\ \citenamefont
  {Glotzer}(2012)}]{jankowski11}%
  \BibitemOpen
  \bibfield  {author} {\bibinfo {author} {\bibfnamefont {E.}~\bibnamefont
  {Jankowski}}\ and\ \bibinfo {author} {\bibfnamefont {S.~C.}\ \bibnamefont
  {Glotzer}},\ }\href@noop {} {\bibfield  {journal} {\bibinfo  {journal} {Soft
  Matter}\ }\textbf {\bibinfo {volume} {8}},\ \bibinfo {pages} {2852} (\bibinfo
  {year} {2012})}\BibitemShut {NoStop}%
\bibitem [{\citenamefont {Ju{\'a}rez}\ and\ \citenamefont
  {Bevan}(2012)}]{bevan12}%
  \BibitemOpen
  \bibfield  {author} {\bibinfo {author} {\bibfnamefont {J.~J.}\ \bibnamefont
  {Ju{\'a}rez}}\ and\ \bibinfo {author} {\bibfnamefont {M.~A.}\ \bibnamefont
  {Bevan}},\ }\href@noop {} {\bibfield  {journal} {\bibinfo  {journal} {Adv.
  Func. Mat.}\ }\textbf {\bibinfo {volume} {22}},\ \bibinfo {pages} {3833}
  (\bibinfo {year} {2012})}\BibitemShut {NoStop}%
\bibitem [{\citenamefont {Tang}\ \emph {et~al.}(2016)\citenamefont {Tang},
  \citenamefont {Rupp}, \citenamefont {Yang}, \citenamefont {Edwards},
  \citenamefont {Grover},\ and\ \citenamefont {Bevan}}]{bevan16}%
  \BibitemOpen
  \bibfield  {author} {\bibinfo {author} {\bibfnamefont {X.}~\bibnamefont
  {Tang}}, \bibinfo {author} {\bibfnamefont {B.}~\bibnamefont {Rupp}}, \bibinfo
  {author} {\bibfnamefont {Y.}~\bibnamefont {Yang}}, \bibinfo {author}
  {\bibfnamefont {T.~D.}\ \bibnamefont {Edwards}}, \bibinfo {author}
  {\bibfnamefont {M.~A.}\ \bibnamefont {Grover}}, \ and\ \bibinfo {author}
  {\bibfnamefont {M.~A.}\ \bibnamefont {Bevan}},\ }\href@noop {} {\bibfield
  {journal} {\bibinfo  {journal} {ACS Nano}\ }\textbf {\bibinfo {volume}
  {10}},\ \bibinfo {pages} {6791} (\bibinfo {year} {2016})}\BibitemShut
  {NoStop}%
\bibitem [{\citenamefont {Lindquist}, \citenamefont {Jadrich},\ and\
  \citenamefont {Truskett}()}]{truskett-arxiv16}%
  \BibitemOpen
  \bibfield  {author} {\bibinfo {author} {\bibfnamefont {B.~A.}\ \bibnamefont
  {Lindquist}}, \bibinfo {author} {\bibfnamefont {R.~B.}\ \bibnamefont
  {Jadrich}}, \ and\ \bibinfo {author} {\bibfnamefont {T.~M.}\ \bibnamefont
  {Truskett}},\ }\href@noop {} {}\bibinfo {note} {Arxiv:1609.00851}\BibitemShut
  {NoStop}%
\bibitem [{\citenamefont {Bar-Haim}\ and\ \citenamefont
  {Klafter}(1998)}]{bar-haim-1998}%
  \BibitemOpen
  \bibfield  {author} {\bibinfo {author} {\bibfnamefont {A.}~\bibnamefont
  {Bar-Haim}}\ and\ \bibinfo {author} {\bibfnamefont {J.}~\bibnamefont
  {Klafter}},\ }\href@noop {} {\bibfield  {journal} {\bibinfo  {journal} {J.
  Chem. Phys.}\ }\textbf {\bibinfo {volume} {109}},\ \bibinfo {pages} {5187}
  (\bibinfo {year} {1998})}\BibitemShut {NoStop}%
\end{thebibliography}%

\end{document}